\documentclass{egu}
\journal{\AG}
\usepackage{amsmath,amsfonts}

\begin{document}

\title{Blob formation and acceleration in the solar wind: role of converging flows and viscosity}
\author{Giovanni Lapenta$^{1,2}$}
\author{Anna Lisa Restante$^{2,3}$}
\affil{$^1$ Centrum voor
Plasma-Astrofysica, Departement Wiskunde,  Katholieke Universiteit
Leuven, Celestijnenlaan 200B - bus 2400, B-3001 Heverlee, Belgi\"e.}
\affil{$^2$ Plasma Theory Group, Theoretical Division, Los Alamos
National Laboratory,  Los Alamos, NM 87545, USA.}
 \affil{$^3$ Solar and Magnetospheric MHD Theory Group, Mathematical Institute,
 North Haugh, University of St Andrews, St Andrews, Fife KY16 9SS,
 Scotland, UK.}
\runningtitle{Blob formation in the solar wind}
\runningauthor{G. Lapenta and A.L. Restante}
\correspondence{Giovanni Lapenta \\(giovanni.lapenta@wis.kuleuven.be)}
\maketitle

\begin{abstract}
The effect of viscosity and of converging flows on the formation of blobs in the slow solar wind is analysed by means of resistive MHD simulations. The regions above coronal streamers where blobs are formed \citep{sheeley} are simulated using a model previously proposed by \citet{einaudijgr}. The result of our investigation is two-fold. First, we demonstrate a new mechanism for enhanced momentum transfer between a forming blob and the fast solar wind surrounding it. The effect is caused by the longer range of the electric field caused by the tearing instability forming the blob. The electric field reaches into the fast solar wind and interacts with it, causing a viscous drag that is global in nature rather than local across fluid layers as it is the case in normal uncharged fluids (like water). Second, the presence of a magnetic cusp at the tip of a coronal helmet streamer causes a converging of the flows on the two sides of the streamer and a direct push of the forming island by the fast solar wind, resulting in a more efficient momentum exchange. 
\end{abstract}

\introduction

The understanding of slow solar wind genesis has progressed considerably in the last few years following the observational discovery of plasma inhomogeneities (called {\em blobs}) formed and expelled from regions above coronal streamers~\citep{sheeley, wang}. The observational evidence has been provided by the Large Angle and Spectrometric Coronagraph (LASCO) instrument on
the Solar and Heliospheric Observatory (SOHO).   Approximately 4
blobs per day were observed in a relatively quiet period for the
solar corona (year 1997, near solar minimum). After formation the blobs are carried away with the ambient slow solar wind. Generally, the plasma blobs move radially outward with their speed increasing from
0-250 km s$^{-1}$ in the region 2-6 ${\rm R}_\odot$ (C2 field of
view) to 200-450 km s$^{-1}$ in the region 3.7-30 ${\rm R}_\odot$
(outer portion of the C3 field of view).

Models of the formation of blobs have focused on mechanisms involving   reconnection between open field
lines and closed field lines \citep{einaudijgr, einaudi,
endeve,fisk, vanaalst, wang, wu}. Through field line reconnection,
plasma from the denser regions within the helmet streamer is
liberated and becomes tied to open field lines and forms the slow
solar wind \citep{einaudi}.

 The present work is particularly centred around the model proposed by \citet{einaudijgr}.  In the region downstream of the cusp in the streamer
belt, \citet{einaudijgr}  approximate the
configuration as a 1D  reversed field configuration with a
1D wake velocity profile. Under these assumptions, they can
reproduce both the acceleration of the slow solar wind and the
formation of blobs, formulating a complete picture of the genesis
of the slow solar wind, in accordance with the observations
\citep{einaudijgr}. The scenario has recently been further
extended~\citep{rappazzo} to include the so called {\em melon seed
effect} due to the diamagnetic force caused by the overall magnetic field radial gradients in the spherical geometry of the Sun~\citep{schmidt} and to include the effects of the cusp magnetic configurations atop the closed field lines in coronal streamers~\citep{lapenta-wind}.

The fundamental conclusion of the model by \citet{einaudijgr} and in successive refinements is that a blob is formed by the tearing instability and that by virtue of the interaction with the fast solar wind the blob is accelerated and ejected, carrying with it the plasma that forms the slow solar wind. 

The fundamental question posed by the present work is: ``what is the basic physical process for the plasma acceleration''? We consider specifically two possibilities. First, that the momentum transfer causing the blob acceleration is due to viscous drag. Second, that the acceleration is due to the insurgence of the tearing instability and its non-linear effects. The acute reader will not find the suspense spoiled by learning that, as usual, Nature turns out not to be black or white. We discover that both mechanisms are present in a interplay mediated by the electric field. Furthermore, the presence of the converging flows at the cusp of the coronal streamer causes a direct momentum transfer by the action of the plasma  flowing against the forming island and pushing  it upward from the solar surface.

The paper is organised as follows. Section 2 summarises the initial configurations considered distinguishing the 1D initial configuration proposed by \citet{einaudijgr} and its 2D extension to include the presence of a magnetic cusp~\citep{lapenta-wind}. Section 3 very briefly provides the methodology for our investigation: the resistive MHD model as implemented in the FLIP3D-MHD code~\citep{brackbill}. Section 4 and 5 present the non-linear evolution, respectively, of the case of the initial 1D reversed field and of the 2D configuration with a magnetic cusp. Conclusions are reached in Section 6.

\section{Initial configuration}

The region where the blobs are formed and presumably reconnection
happens has been localised thanks to the LASCO observations. The
blobs are formed between 2-3 ${\rm R}_\odot$ (the cusp region) and
are accelerated afterwards, reaching terminal speed 250-450 km
s$^{-1}$ at 20-30 ${\rm R}_\odot$. The sonic point where the fast
solar wind reaches sonic speeds is estimated at 5-6 ${\rm R}_\odot$,
the Alfven point is further outwards. In the region of blob
formation the fast wind is slightly subsonic and subalfvenic. In the present paper we consider the region where the plasma blobs
and the slow solar wind originate. We consider the region near the
cusp where we assume the plasma to be near sonic but subalfvenic.

We consider two types of configurations previously proposed for the regions where blobs are observed to form. The first is 1D and is a simple force-free extension of the Harris equilibrium~\citep{einaudi}. The second, is a more complex 2D equilibrium representative of a Helmet streamer~\citep{lapenta-wind}.
In both cases, the central current sheet represents the area above the tip of a helmet streamer where the blobs have been observed to form in the LASCO images. To model the presence of the fast solar wind emanating from the coronal holes, the initial equilibrium includes the presence of a flow along the field lines. The flow is chosen to be zero along the field lines directly above the streamer tip and to increase rapidly away from it, to reach an asymptotic value corresponding to the fast solar wind. The details of the initial configurations have been presented in previous publications and are summarised below for each of the two cases.

We use Cartesian co-ordinates where $y$ and $z$ are parallel to
the Sun's surface (with $z$ along the meridians  and $y$ along the
equator) and $x$ is orthogonal to and pointed away from the Sun's
surface.

\subsection{1D force-free equilibrium}

We use the 1D model proposed by \citet{einaudi}. 
The magnetic field is force free (i.e. the pressure, temperature and density are uniform) and is given by
\begin{equation}
\begin{array}{c}
B_{0x} =B_0 \tanh((z-z_0)/L) \\
B_{0y} =B_0 {\rm sech}((z-z_0)/L)
\end{array}
\end{equation}
where $L$ is the length scale of the variation of $B$. A plasma flow  with a wake
profile is chosen:
\begin{equation}
u_{0x} (z) =u_0 {\rm sech}((z-z_0)/L);
\end{equation}
The flow profile is chosen according to the model of \citet{einaudi} and represents the fast solar wind propagating along the open field lines, while a stationary plasma lays above the helmet streamer. As noted in \citet{einaudi}, the scale-length of variation of the flow shear might differ significantly from the length scale of variation of the magnetic field. However, for simplicity we assume them to be the same. In the present paper we choose the particular case $u_0/v_A=2/3$, where the Alfv\'en speed $v_A$ is defined with $B_0$. 

 Periodic boundary conditions are applied in the $x$ direction and free-slip boundary conditions are applied in $z$. The boundaries in $z$ allow no outflow of plasma and form effectively a closed plasma channel.  A small perturbation is added to the system to stimulate the growth of any instability present. The perturbation is added to the $y$ component of the vector potential as:
\begin{equation}
 \delta A_y = \epsilon \sin(k_x x) sin(k_z z)
\end{equation}
with $k_x=2 \pi/L_x$, $k_z=\pi/L_z$ and $\epsilon=B_0/15$.

\subsection{2D Helmet-streamer}
 The second initial configuration is based on a
recent work by \citet{schindler} where a
magnetic field configuration with a current sheet typical of a
helmet streamer is computed analytically. The equilibrium is rigourously exact only for infinitely stretched helmet streamers, but it can be used even for realistic aspect ratios as a good approximation.

\begin{figure}
 \centering
 \hskip -1cm
\includegraphics[width=80mm]{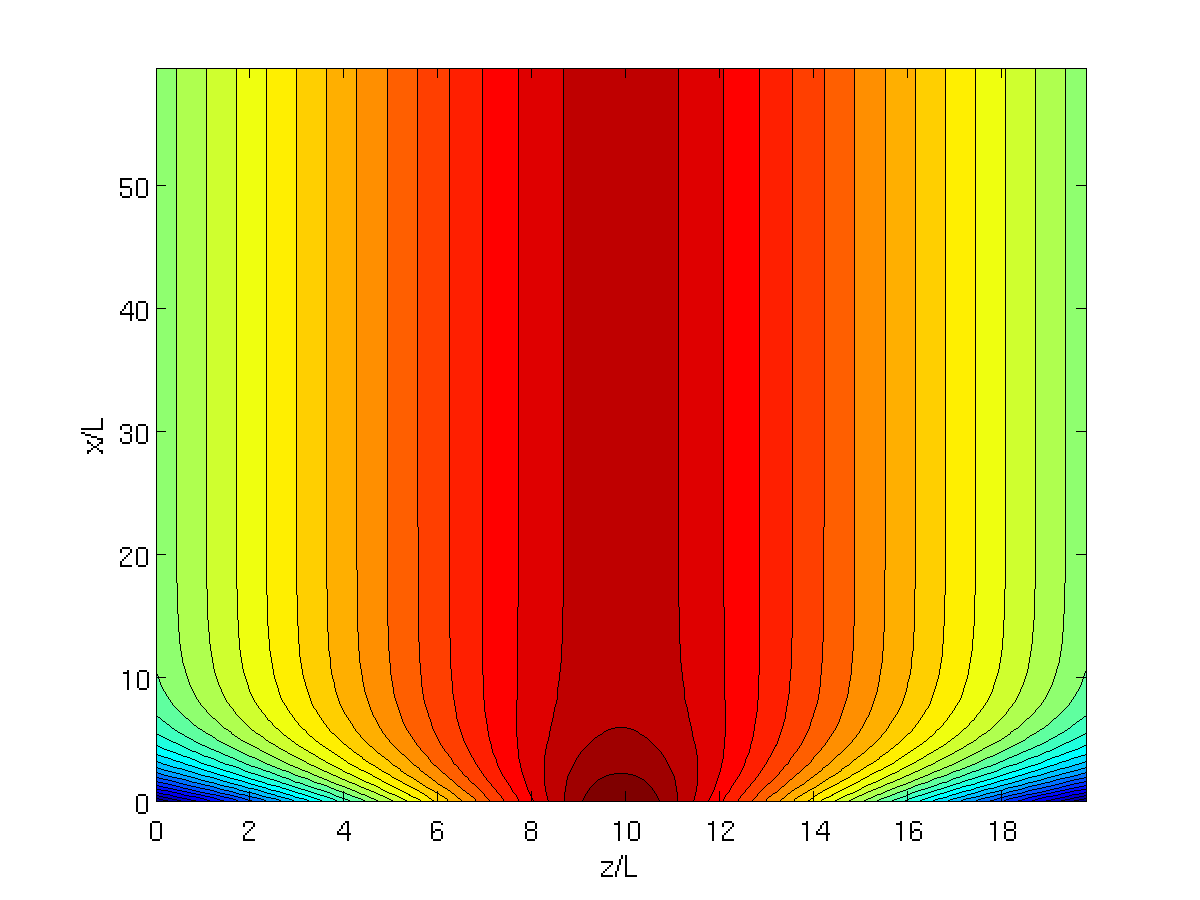}
\caption{Initial Helmet Streamer equilibrium. Note that the vertical and horizontal axis are not on the
same scale, the vertical axis in compressed.} \label{initial}
\end{figure}

In the co-ordinate system used here, the current is initially directed along $y$  and the initial magnetic field is in the $(x,z)$ plane (see Fig. ~\ref{initial}), with $x$ being the direction away from the Sun's surface.

The initial equilibrium is 2D and is independent of the $y$ co-ordinate.
The initial configuration is characterised by the $y$ component of the vector potential:
\begin{equation}
A_y(x,z)=-\frac{2}{c} \log \left\{ \cosh \left [ c \sqrt{\frac{p_0}{2}(z-z_0)}\right] \right\}
+\frac{1}{c} \log \frac{p_0}{k} \label{initA}
\end{equation}
where $c$, $z_0$ and $k$ are constants and $p_0(x)=s_1 e^{-s_2 x}
+s_3$ The initial magnetic configuration has a current:
\begin{equation}
j=\frac{p_0c}{\cosh^2 \left( \sqrt{\frac{p_0}{2}} c (z-z_0) \right)}\label{initj}
\end{equation}
 and is held in pressure equilibrium
by a pressure profile:
\begin{equation}
p=\frac{p_0}{\cosh^2 \left( \sqrt{\frac{p_0}{2}} c (z-z_0) \right)} +p_B\label{initp}
\end{equation}
The  separatrices between open and close field lines are
positioned symmetrically at equal distances from the centre $z_0$:
\begin{equation}
z_{sep}= \frac{1}{c} \sqrt{\frac{2}{p_0}}{\rm arctanh}\left
(\sqrt{\frac{p_0-\varsigma}{p_0}} \right)
\end{equation}
where $\varsigma=k e^{-cA_s}$.
 In the simulations shown below we choose the parameters as \citep{wiegelmann, schindler, lapenta-solar}
 $s_1=.8$, $s_2=4$, $s_3=.2$, $c=15$, $A_s=.1073$  and
$k=1$, $z_0=1$. Following \citet{wiegelmann}, a  small
($20\%$) background plasma pressure $p_B$ is added to simplify the
numerical treatment. The initial configuration is assumed to have
uniform plasma density and the specific internal energy profile
can be derived from eq.(\ref{initp}):
\begin{equation}
I=\frac{ p}{\rho(\gamma-1)} \label{inite}
\end{equation}

Additionally, we add an initial wake velocity field:
\begin{equation}
{\bf u}(z)= 
 \left[1-{\rm sech}
\left(\frac{|z-z_0|-z_{sep}}{L_v}\right) \right] e^{-\frac{|z-z_0|-z_{sep}}{2L_v}}
\end{equation}
where ${\bf \widehat{b}}$ is the unit vector in the direction of
the magnetic field, where $L_v$ is the scale-size of variation of the velocity field. 
The flow is present only outside of the two
separatrices, where the field lines are open and is zero inside,
where the field lines are closed.

At the bottom boundary ($x=0$), the field lines are tied and the
flow and the plasma are continuously resupplied with the same
value used for the initial condition to simulate the fast wind
ejected from the coronal holes. At all other boundaries, open
conditions are applied to allow the outflow of plasma.

Unlike the previous case, no initial perturbation is applied since the error in
the approximation of infinite stretching typical of the
derivations leading to eq.~(\ref{initA}) is in itself a sufficient initial
perturbation. 

 In the present 2D equilibrium the equilibrium vary
along $z$ and  the typical thickness of the layer
downstream of the cusp, $L=L_z/20$. As in the 1D case, the velocity is assumed to have the same length scale as the magnetic field: $L_v=L$.

\section{Simulation approach}
The simulations for the present work are conducted with the FLIP3D-MHD code~\citep{brackbill} based on the viscous-resistive MHD model~\citep{goossens}, comprising a
mass continuity equation,
\begin{equation}\label{B_eq_1}
\dfrac{d\rho}{dt}+\rho \nabla \cdot \mathbf{u}=0
\end{equation}
Faraday's and Ampere's laws,
\begin{align}\label{B_eq_2}
\dfrac{d}{dt} \left[ \dfrac{\mathbf{B}}{\rho}\right] &=
\dfrac{\mathbf{B}}{\rho} \cdot \nabla \mathbf{u} - \dfrac{1}{\rho}
\left[ \nabla \times \eta J\right]\\ \nonumber \\ \nonumber 
\mu_0 J &= \nabla \times \mathbf{B}
\end{align}
Ohm's law,
\begin{equation}
 \mathbf{E}=\eta\mathbf{J}-\mathbf{u}\times \mathbf{B}
\end{equation}
a momentum equation,
\begin{equation}\label{B_eq_3}
\rho \dfrac{d\mathbf{u}}{dt}= - \nabla \left[ p+
\dfrac{\mathbf{B}^2}{2 \mu_0}\right]+ \left[ \nabla \cdot
\dfrac{\mathbf{B}\mathbf{B}}{\mu_0}\right] + \nabla \lambda \rho
\nabla \cdot \mathbf{u}- \nabla \cdot \nu \rho \Pi
\end{equation}
and an energy equation,
\begin{equation}\label{B_eq_4}
\rho \dfrac{dI}{dt}= - p \nabla \cdot \mathbf{u}+ \lambda \rho
(\nabla \cdot \mathbf{u})^2 + \nu \rho (\Pi \cdot \Pi) + \eta
(\mathbf{J}\cdot\mathbf{J})
\end{equation}
where $\rho$ is the mass density, $\mathbf{B}$ is the magnetic field
intensity, $\mathbf{J}$ is the current density, $c$ is the speed of
light, $\mu_0$ is the magnetic permeability of vacuum, 
$\mathbf{u}$ is the fluid velocity, $I$ is the specific
internal energy, and $p$ is the fluid pressure. The symmetric
rate-of-strain tensor, $\Pi$, is defined in the usual way~\citep{braginskii},
\begin{equation}\label{B_eq_4a}
\Pi= \dfrac{1}{2} [\nabla \mathbf{u} + \nabla \mathbf{u}^T]
\end{equation}
The pressure is given by the equation of state, $ p = (\gamma-1) \rho I$ with $\gamma=5/3$. The
transport coefficients are the kinematic shear viscosity $\nu$, the
kinematic bulk viscosity $\lambda$, and the resistivity
$\eta$. The solenoidal condition on $\mathbf{B}$,
\begin{equation}\label{B_eq_5}
\nabla \cdot \mathbf{B}=0
\end{equation}
is imposed throughout the evolution.

The system under investigation is studied in a two-dimensional plane
($x$, $z$) using FLIP-MHD, a resistive MHD code described in
\citet{brackbill}.   The system size in each direction is
labelled $L_x$,  $L_z$  and a grid made of $300 \times 600$ Lagrangian markers 
is used in each direction arrayed initially in a $3 \times 3$ uniform formation 
in each cell of a $100 \times 200$ grid. The units of the simulations are normalised to the magnetic (and velocity ) scale length $L$ and to the Alfv\'en time $\tau_A=L/v_A$. The Lundquist number~\citep{poedts} is computed as $S=\mu_0 v_A L/\eta$ and the Reynolds number as: $R_v=u_0\rho L/\nu$ (using the initial uniform density and the  fast solar wind speed). We consider here for simplicity only shear viscosity ($\lambda=0$).

\section{Non-linear evolution of the 1D initial equilibrium}

The initial state is characterised by both a flow shear and a magnetic shear. In principle, a number of instabilities could be present~\citep{biskamp,bettarini}. However, for the choice of parameters considered only two types of instabilities are present.

As shown in previously published results \citep{einaudi,lapenta-wind}, the resistive tearing instability is present and grows at a rate determined by the resistivity of the system.  The resistive tearing mode leads to the formation of a magnetic island that mixes outer and inner field lines. In the process, the outer solar wind can transfer its momentum to the inner plasma and produce a net motion of the tearing island. This process is clearly reminiscent of the observed formation of blobs and their subsequent acceleration away from the Sun.  

However,  a second instability is present. Even in absence of resistive MHD modes, the viscosity is itself capable of altering the initial profile. Indeed, a viscous flow would tend to flatten any flow profile in time by virtue of direct viscous drag alone. The inner layer that is initially motionless would progressively acquire speed \citep{landau}. In the corona, the viscosity is very low unless anomalous effects become important. However, in any realistically feasible simulation, numerical viscosity has to be present to avoid the formation of small scale structures that cannot be resolved and would lead to numerical instabilities.

One aim of the present work is to elucidate the relative role of this second process relative to the first. The question is:  is the slow solar wind acceleration observed in the simulations due to the tearing instability or to the viscous drag?  And how does it is scale with viscosity and resistivity,

\subsection{Formation of the slow solar wind and of plasma blobs}

To exemplify the typical evolution observed in the simulations of the 1D initial state, we report the results of a simulation  for the
particular case with Lundquist number  $S=10^4$ and viscous Reynolds number $R_v=10^4$.

\begin{figure*}[t]
\vspace*{2mm}
\begin{tabular}{cc}
A) & B) \\
\includegraphics[width=8
cm]{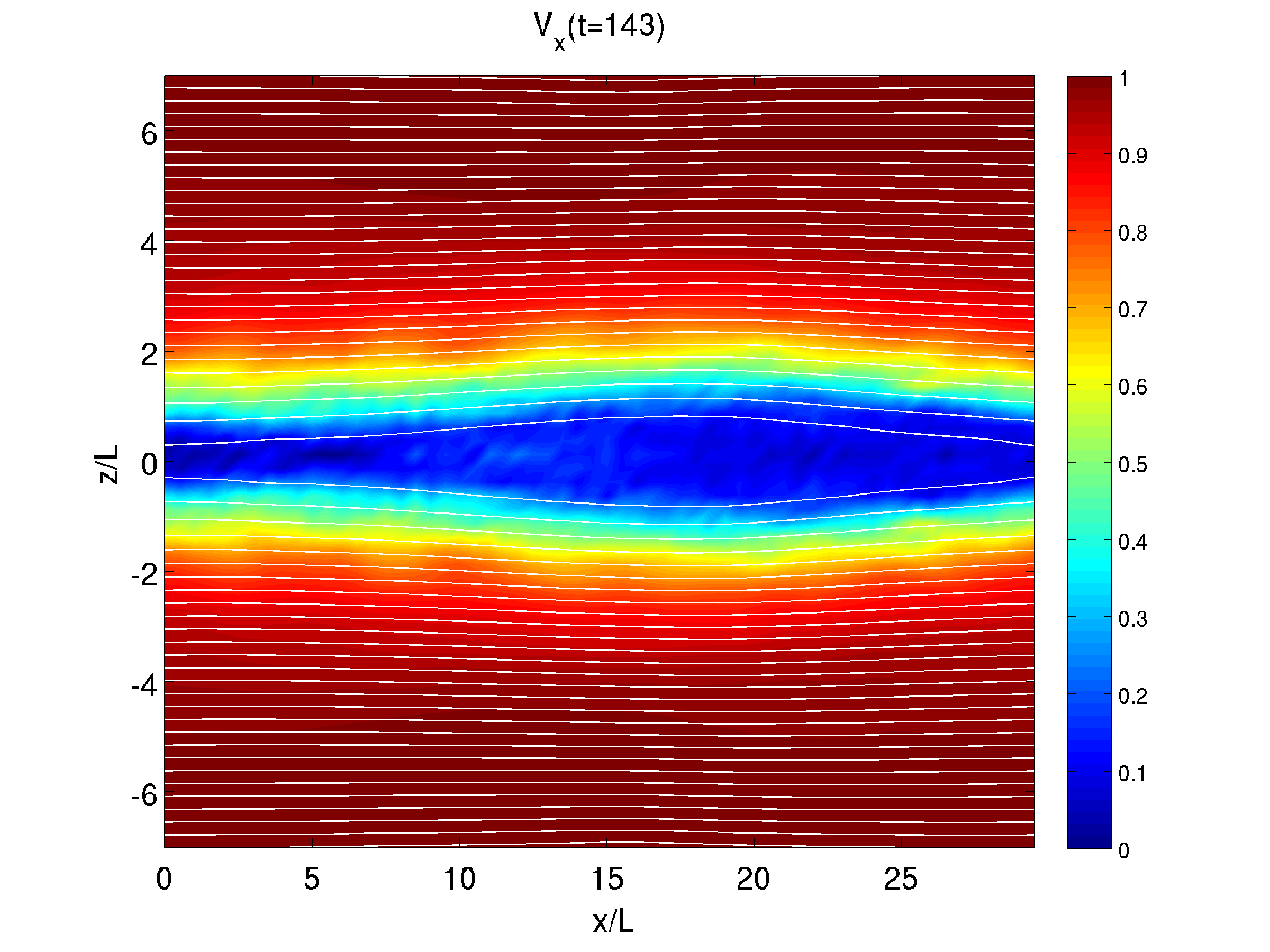} &
\includegraphics[width=8
cm]{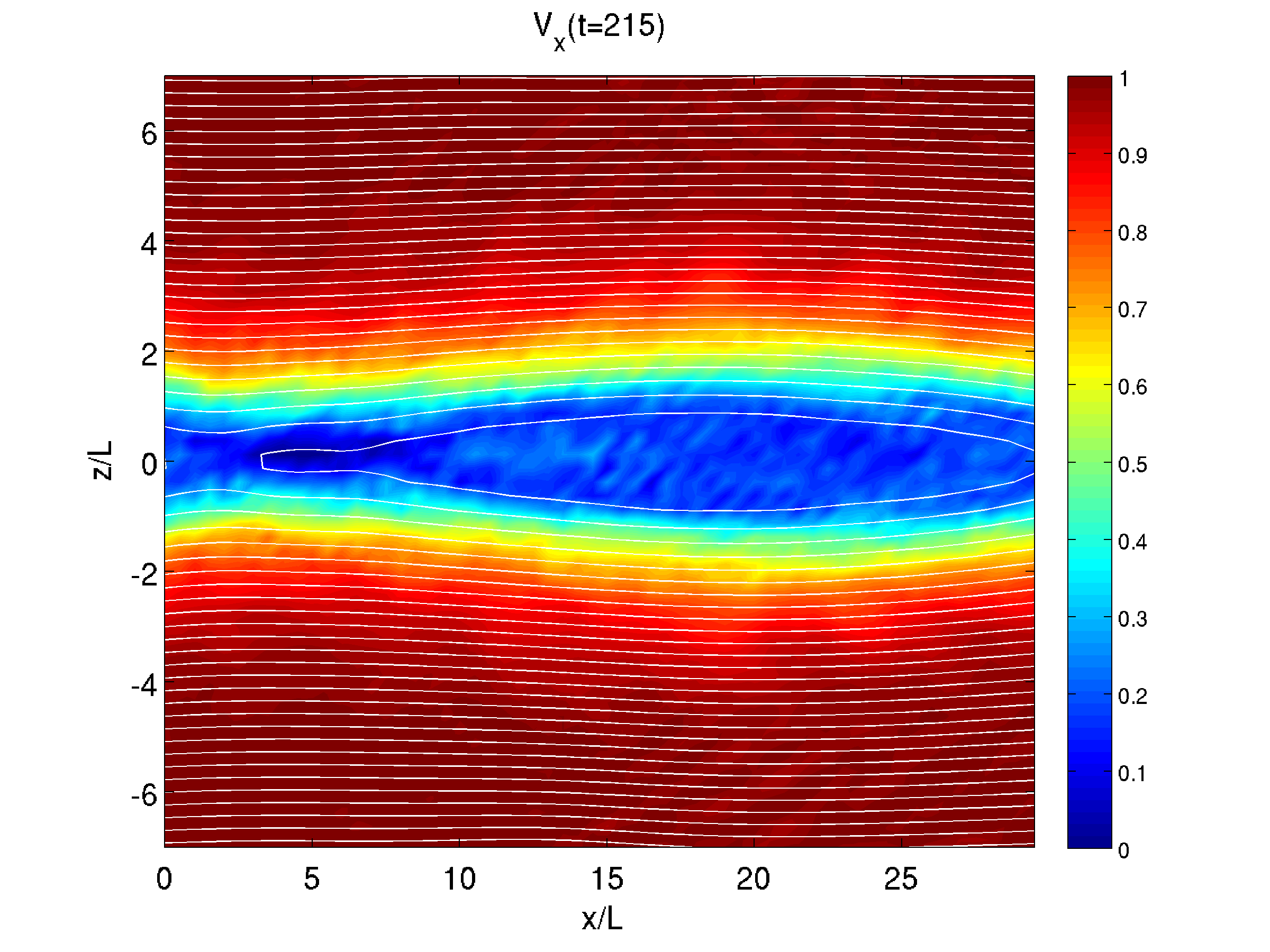}\\
C) & D) \\
\includegraphics[width=8
cm]{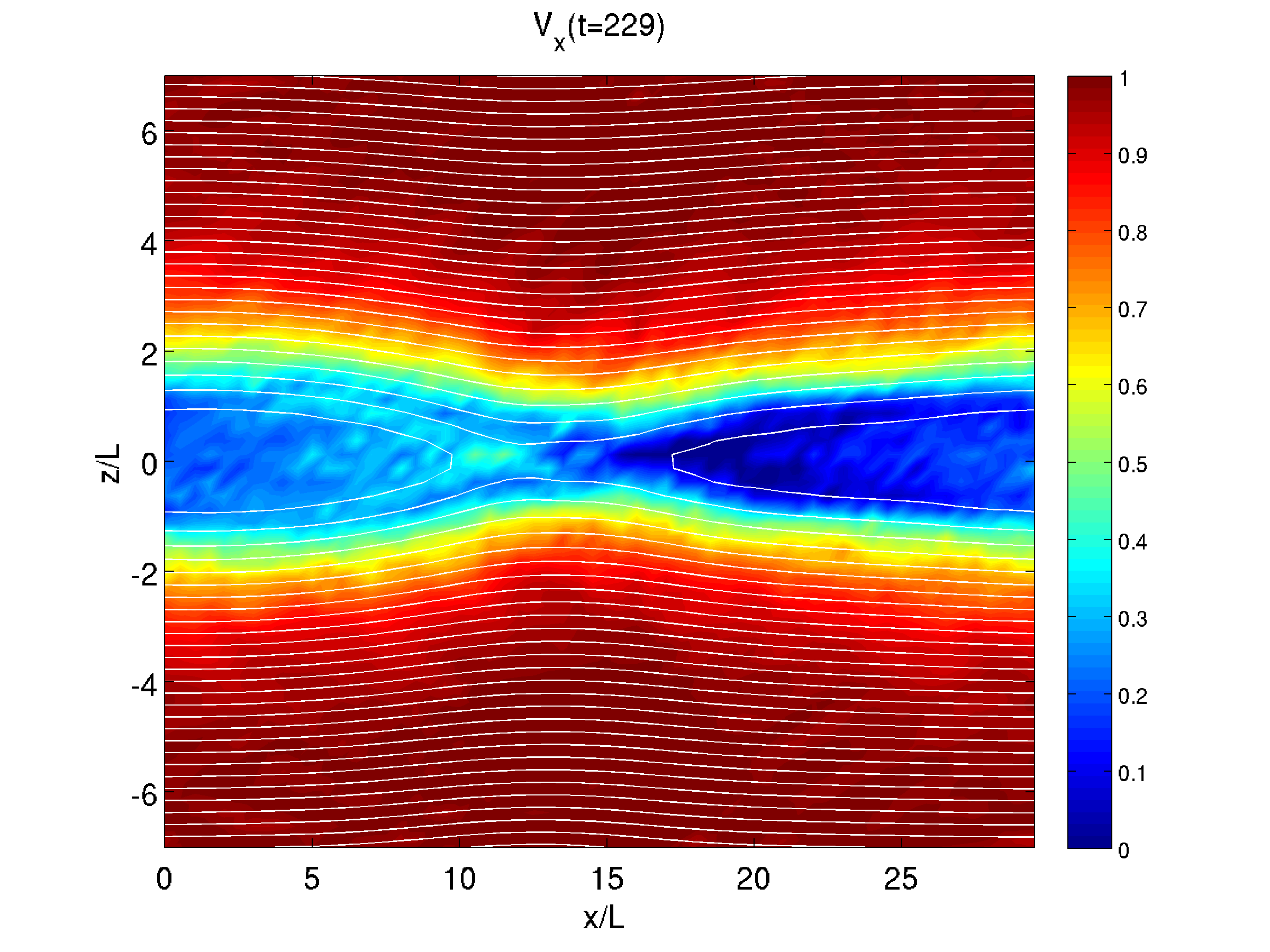} &
\includegraphics[width=8
cm]{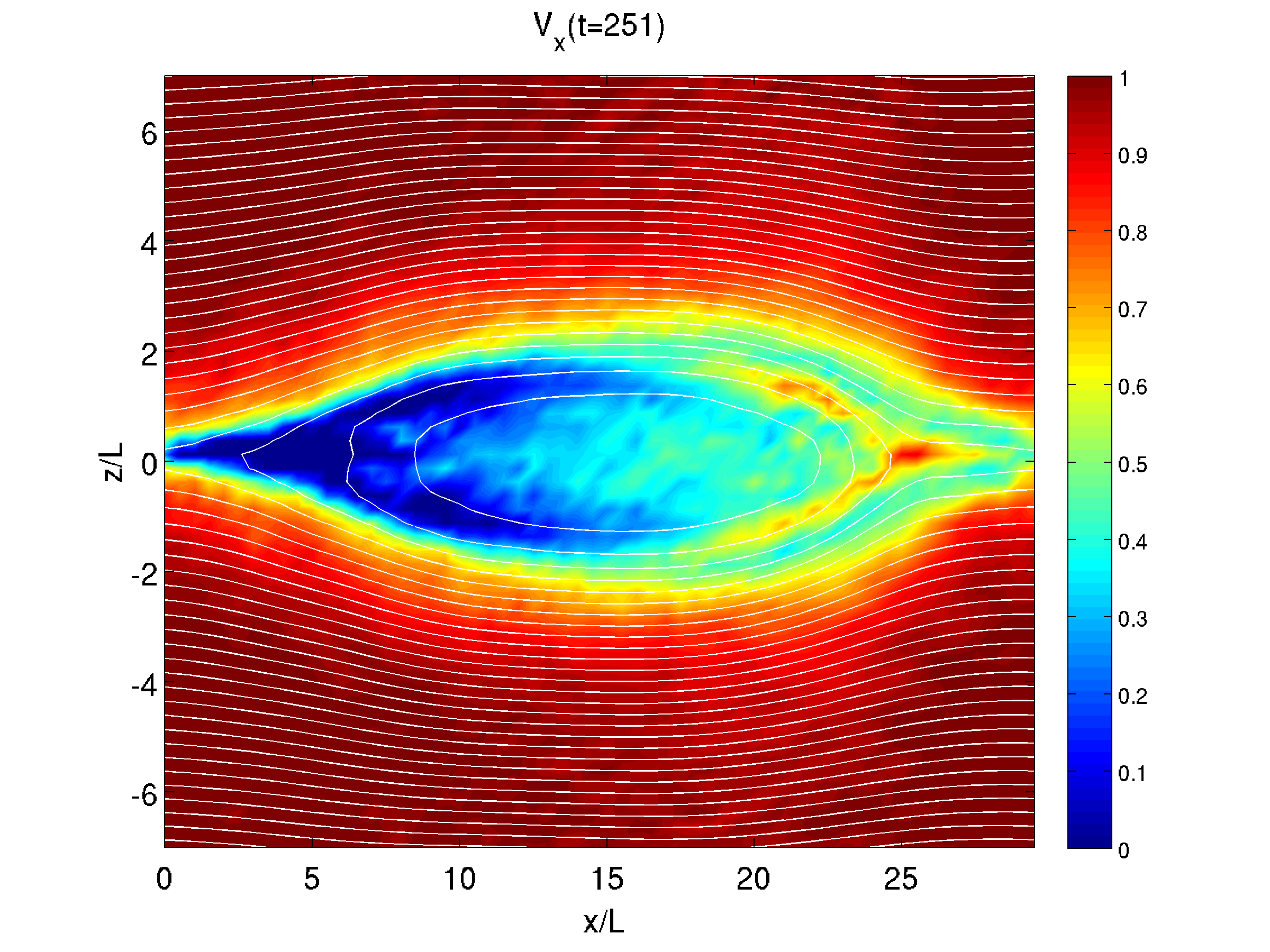}
\end{tabular}
\caption{Non-linear evolution of the initial force free reversed field equilibrium. Velocity component (false colour) away from the Sun and magnetic field lines (white) are shown at four different times (A: $t/\tau_A=143$, B: $t/\tau_A=215$,
C: $t/\tau_A=229$, D: $t/\tau_A=251$) in a run with $S=10^4$ and $R_v=10^4$.} \label{it15}
\end{figure*}

The initial phase of the evolution sees the formation of a magnetic island that perturbs the initial topology of the system and the initial flow profile. Figure \ref{it15} shows the velocity away from the Sun in false colour, superimposed on the magnetic filed lines, at four times showing different phases of island formation. The island grows and moves away from its original formation site. Comparing the different times shown in Fig.~\ref{it15} demonstrates that the island moves along with the flow. Note that in the simulations, the frame of reference is chosen stationary with the plasma flow in the flanks, so the motion of the island seen in the figure is with respect to the fast solar wind.

\begin{figure}[t]
\vspace*{2mm}
\begin{center}
A)\\
\includegraphics[width =8 cm ]{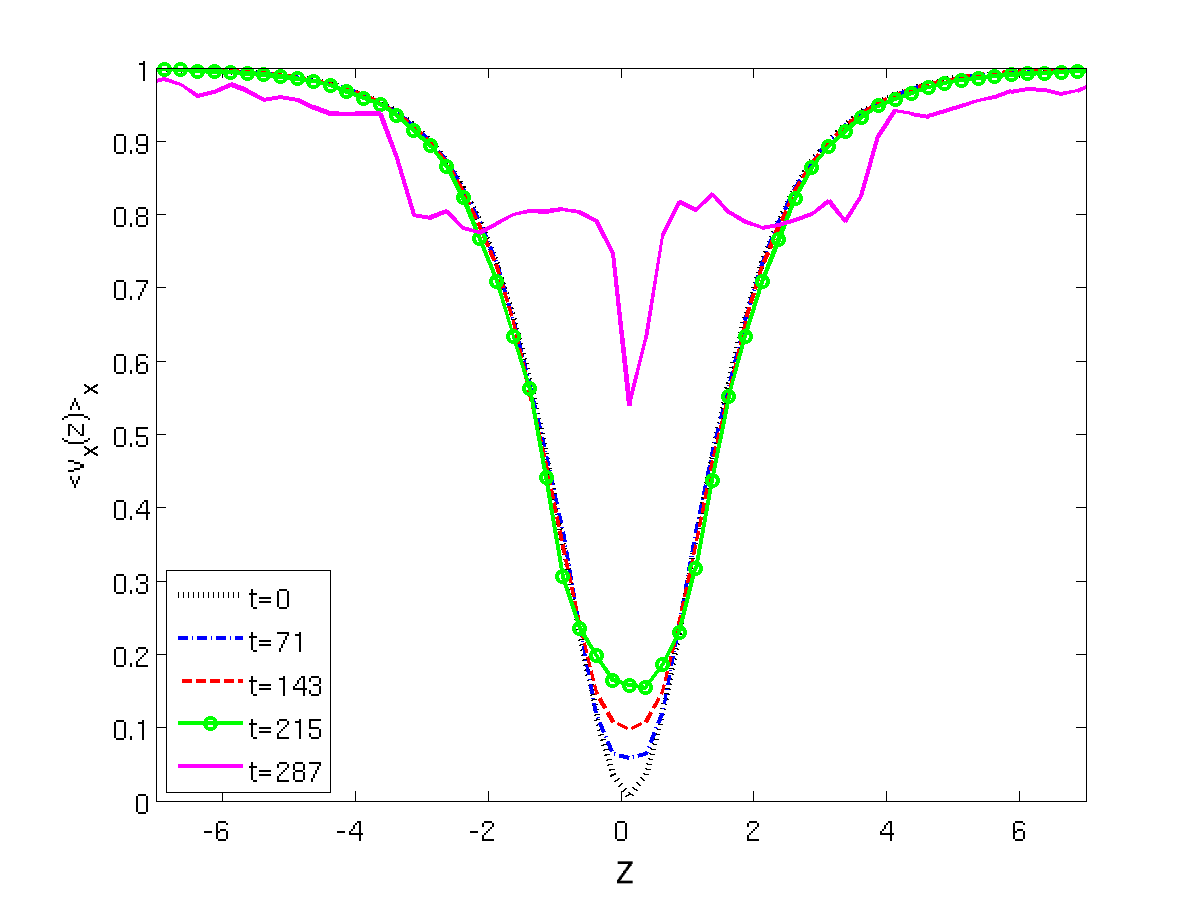}\\
B)\\
\includegraphics[width =8 cm ]{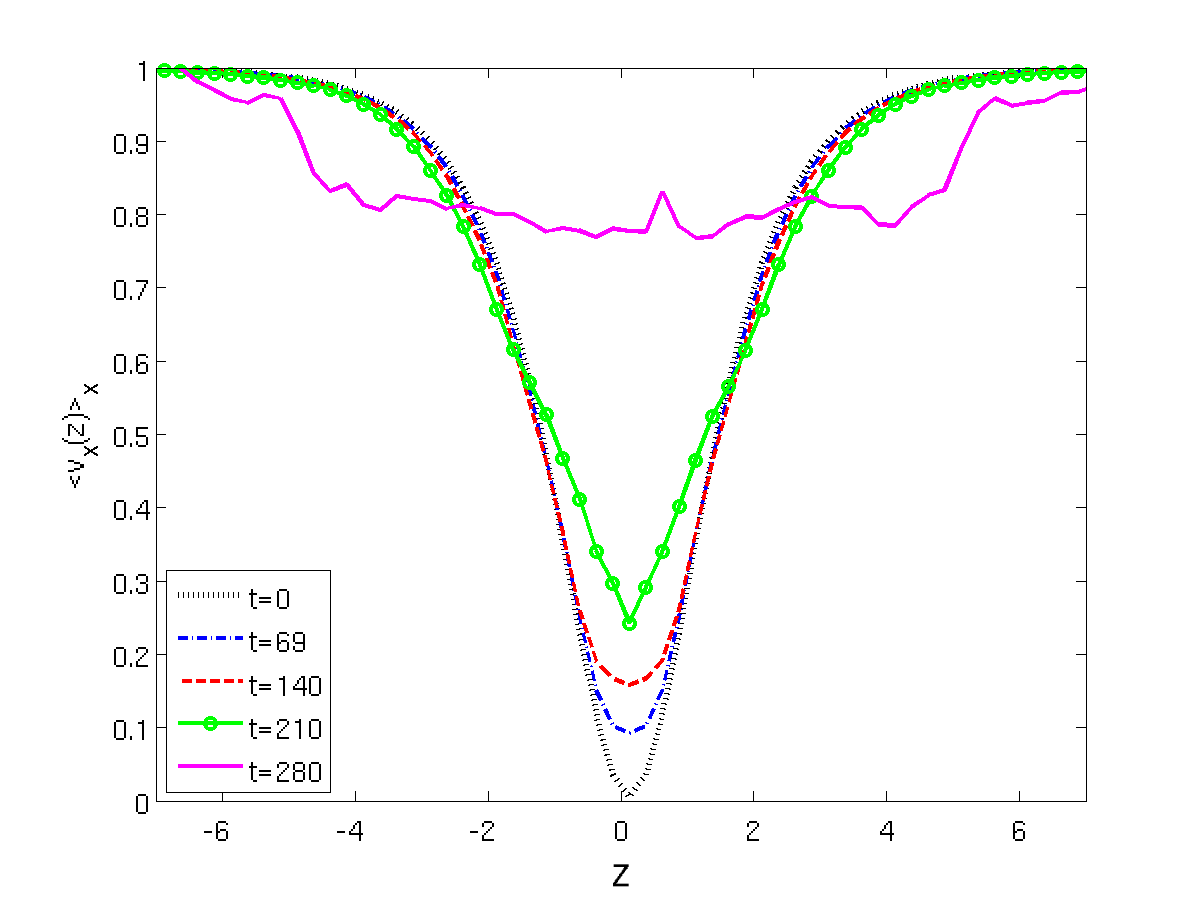}
\end{center}
\caption{Evolution of the average velocity profile $<v_x(z)>_x$ (normalised to the asymptotic speed $u_0$) for two runs starting from the  initial force free reversed field equilibrium with the same resistivity ($S=10^4$) but different viscosity (panel A with $R_v=10^4$, panel B $R_v=10^3$).}\label{vel_r10_mu100}
\end{figure}

\begin{figure}[t]
\vspace*{2mm}
\begin{tabular}{c}
A) \\ 
\includegraphics[width =8 cm ]{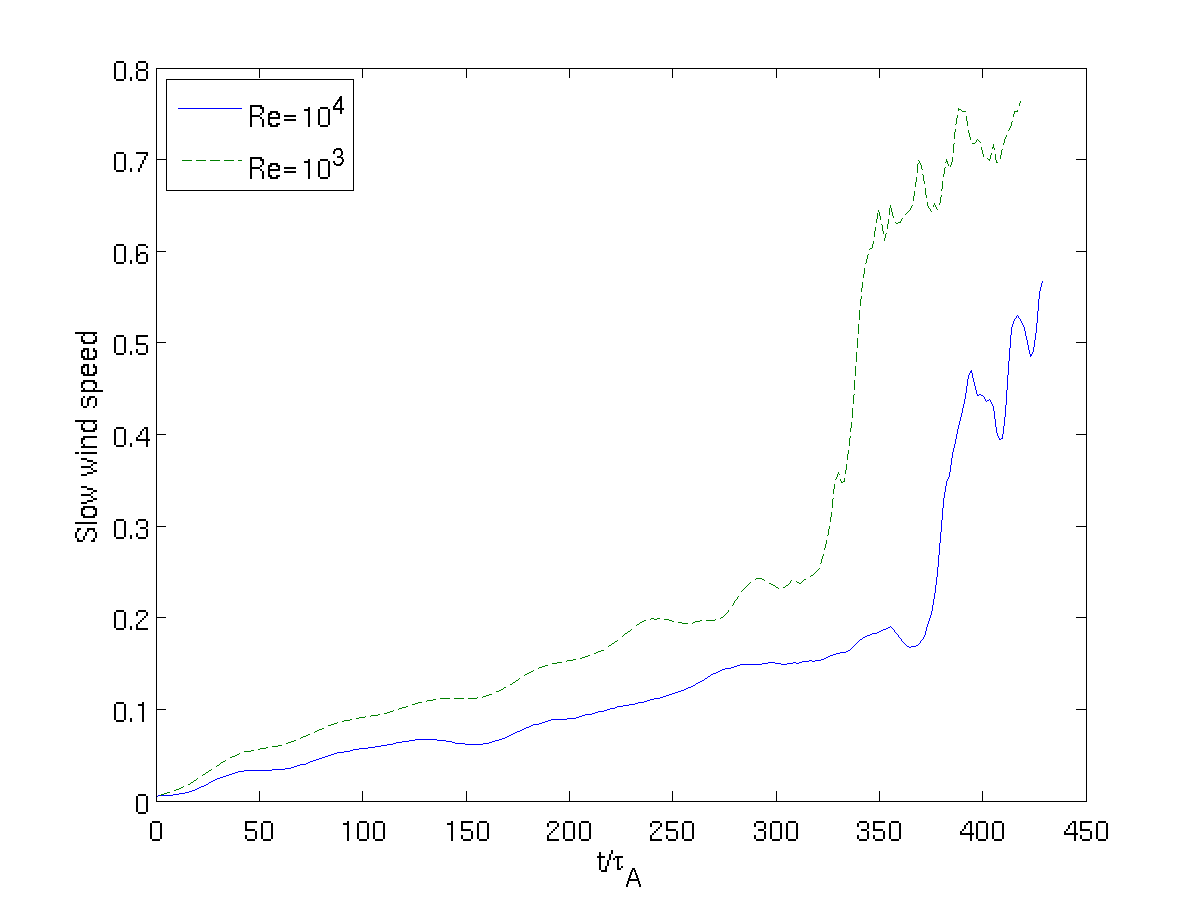} \\B) \\
\includegraphics[width =8 cm ]{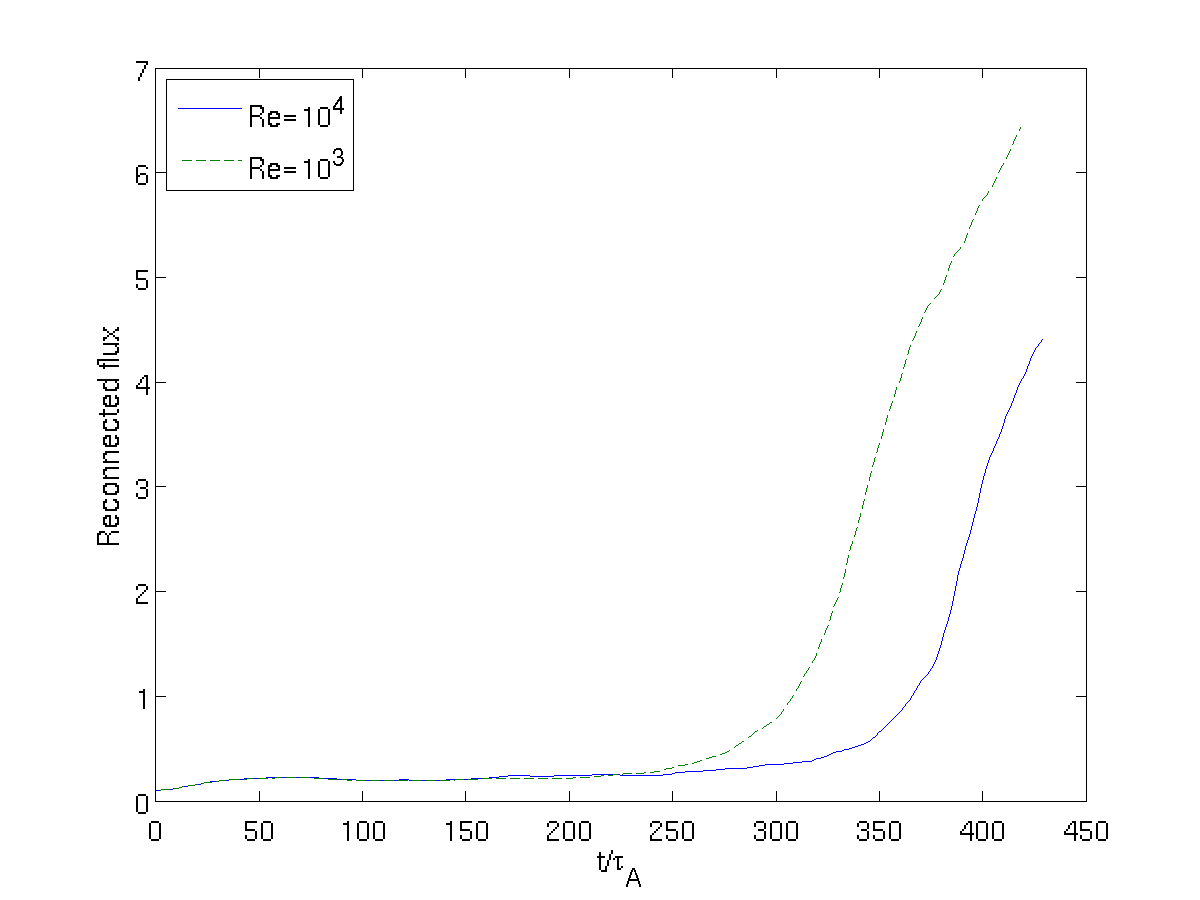}
\end{tabular}
\caption{Evolution of the peak of the average velocity (normalized to the asymptotic speed $u_0$)  $\max_z(<u_x(z)>_x)$ (A) and of the  reconnected flux (B) for two runs  starting from the  initial force free reversed field equilibrium with the same resistivity ($S=10^4$) but different viscosity (panel A with $R_v=10^4$, panel B $R_v=10^3$).}\label{windvst}
\end{figure}

The flow profile at different times is shown in Fig.~\ref{vel_r10_mu100}. The momentum transfer between the flanks and the centre is made evident by the progressive flattening of the profile. The central plasma is progressively accelerated and the speed differential between the centre and the flanks diminishes with time. The temporal evolution of the speed of the central region of the plasma is shown in Fig.~\ref{windvst}-a. Two cases are shown, besides the simulation considered above, another with higher viscosity ($R_v=10^3$) but equal resistivity is also reported. The process of acceleration progresses in two steps.  After a  progressive and continuous phase of acceleration of the initially stagnant plasma, a sudden acceleration develops when the island becomes so big as to feel the effect of the boundary conditions. AS shown below, this last phase is only present in the low resistivity cases and is it is due to a transition from a X-point reconnection process to a Y-point reconnection process with a progressively more elongated leg~\cite{biskamp-recon} becoming progressively more turbulent and undergoing secondary island formation. These features of the reconnection mechanism are well known and will not be addressed here again, the interested reader is referred to the excellent textbook by \citet{biskamp-recon}.

The process of island formation is further shown in Fig.~\ref{windvst}-b where for the same two simulations, the reconnected flux is shown. The onset of the faster reconnection phase also is similarly due to the boundary conditions chocking the flow which turns inward toward the reconnection site at the x-point formed by the island driving faster reconnection. Again this process is an artefact of the boundary conditions. 

Figure \ref{windvst} shows a comparison of two cases with different viscosity and demonstrates that indeed viscosity has a strong impact on both island formation and slow solar wind acceleration. In the next paragraph this effect is further analysed. 

\subsection{Role of Viscosity and Resistivity}

The behaviour displayed above is in agreement with previously published results~\citep{einaudi,lapenta-wind} and it is reproduced primarily to provide the reader with a summary of the typical evolution. Below we investigate the main point of the present effort, namely what are the causes of the observed behaviour displayed in Fig.~\ref{vel_r10_mu100}. Both the tearing instability and viscous stresses can be the cause: to determine which dominates, we change independently viscosity and resistivity. The tearing instability is proportional to resistivity, viscous drag is proportional to viscosity. We need to determine whether the momentum transfer is proportional to viscosity or resistivity.

\begin{figure}
\begin{tabular}{cc}
A) $S=10^4$ & B) $S =2 \cdot 10^3$ \\
\includegraphics[width=4 cm ]{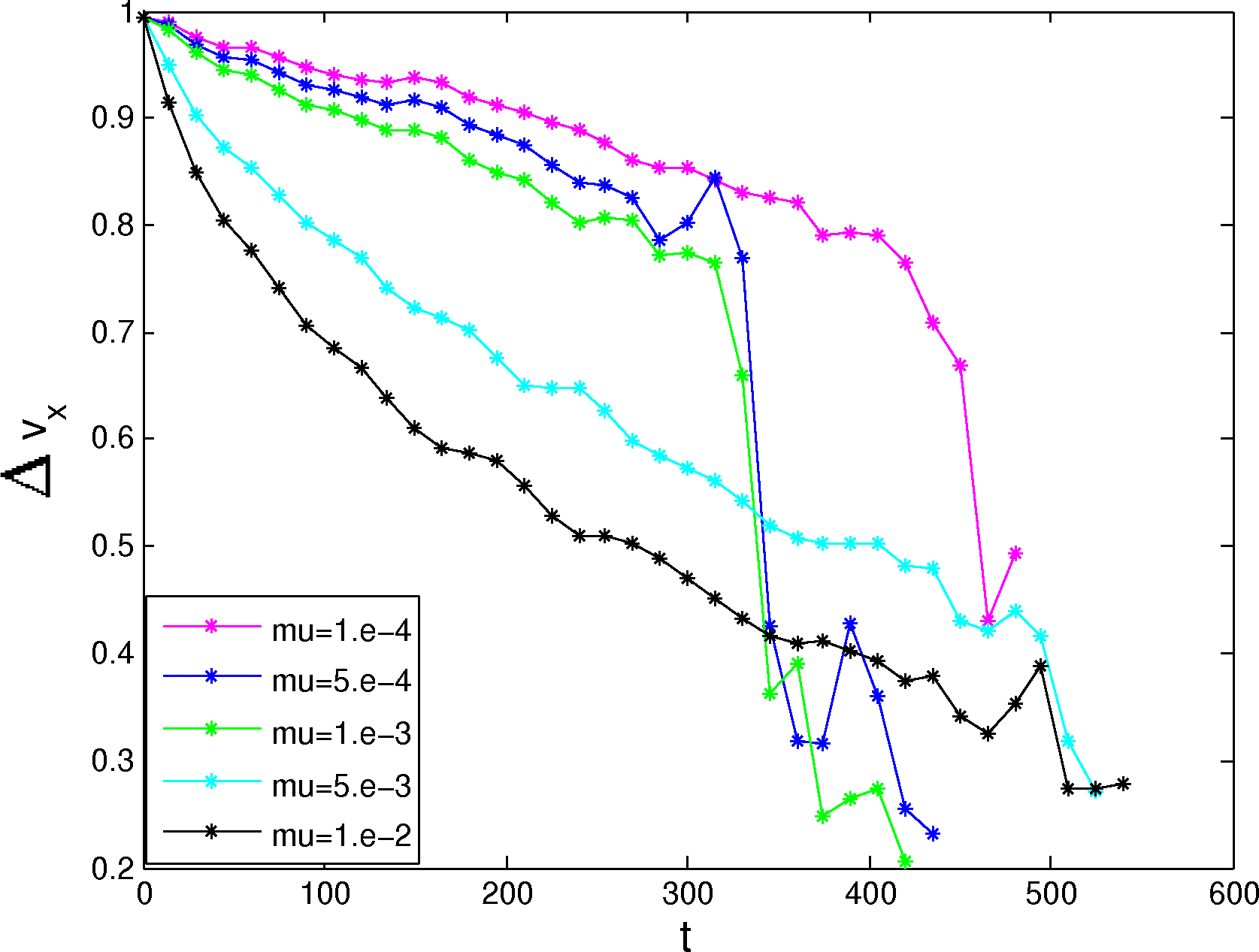} &
\includegraphics[width =4 cm ]{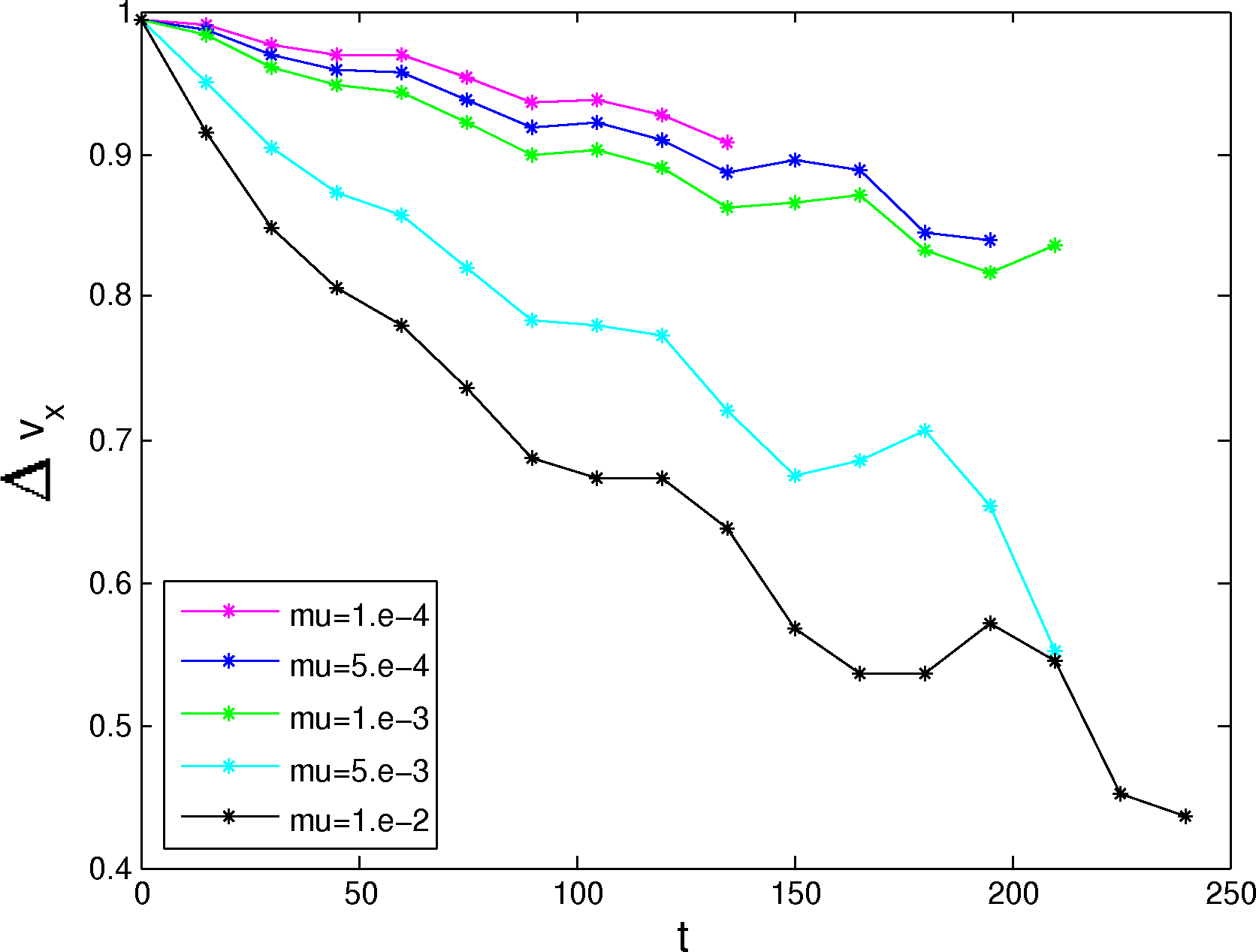}\\
C) $S=10^3 $ & D) $S= 2 \cdot 10^2$ \\
\includegraphics[width =4 cm ]{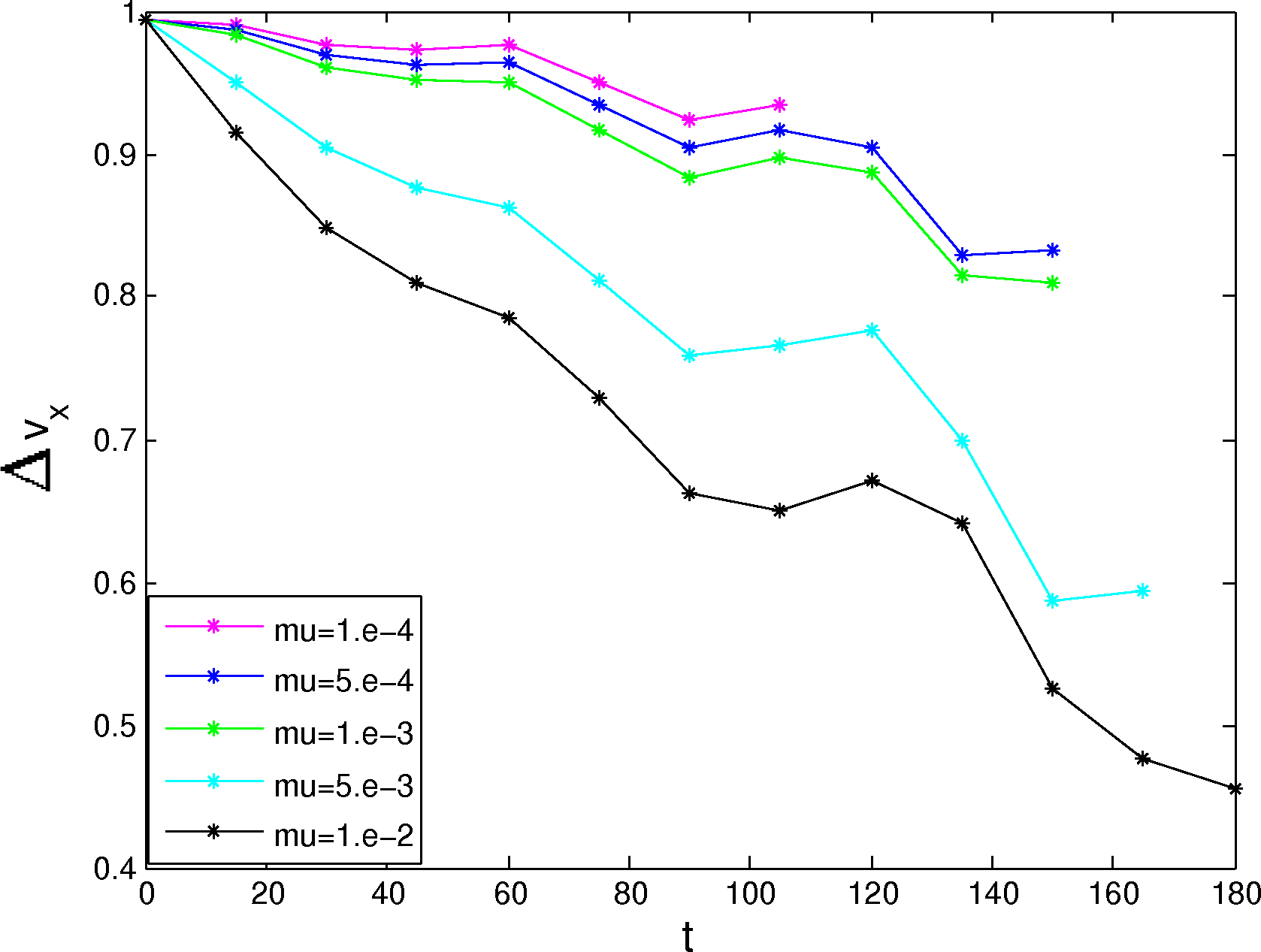}
&\includegraphics[width =4 cm ]{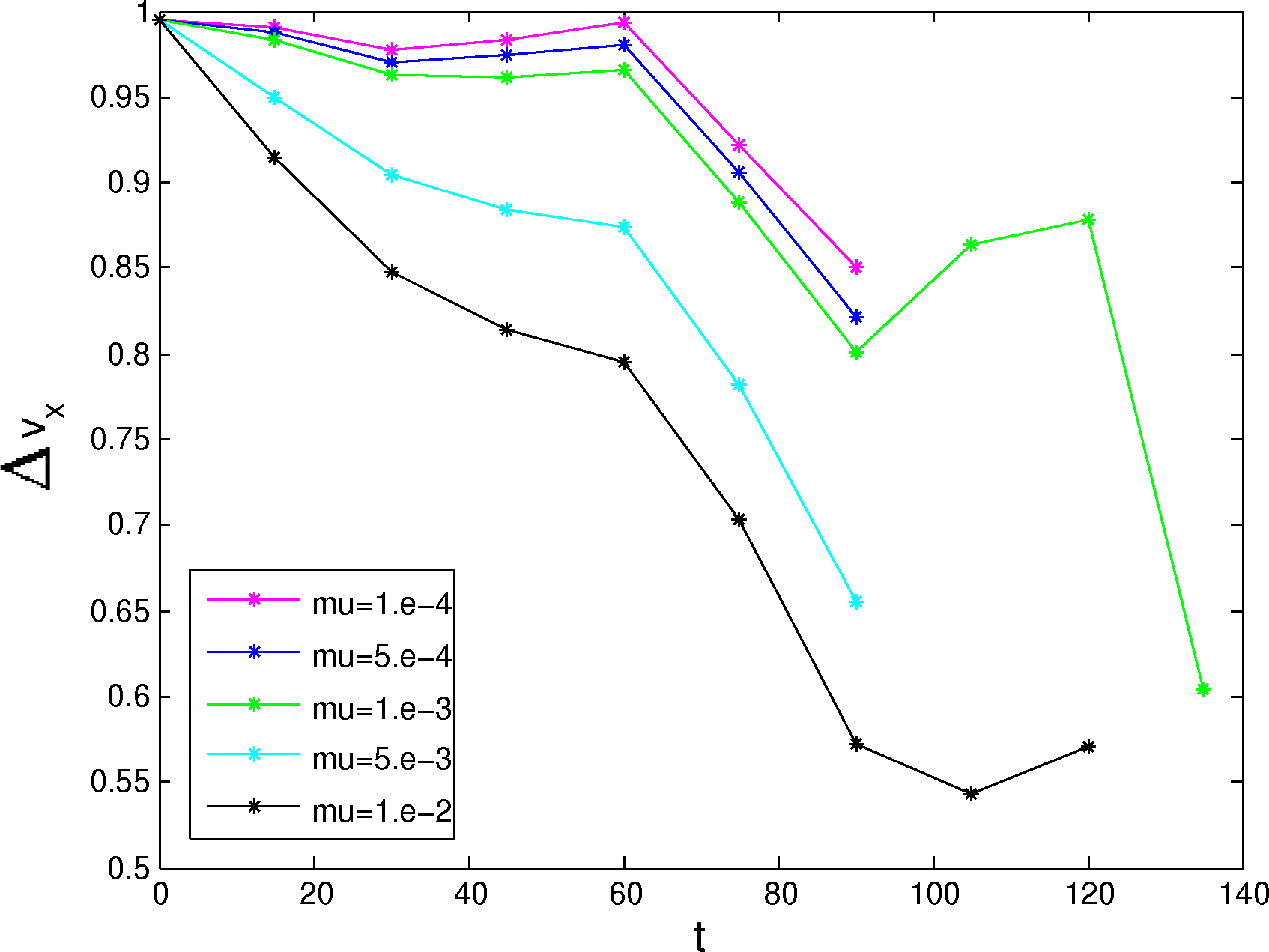}
\end{tabular}
\caption{Evolution of the difference $\Delta V_x$ between the maximum (fast solar wind) and minimum (slow solar wind) in the profile of the average vertical velocity (normalized to the asymptotic speed $u_0$)  $\max_z(<u_x(z)>_x)$. Several runs are shown  starting from the same initial force free reversed field equilibrium, each panel corresponds to a different resistivity: A) $S=10^4$, B) $S=2 \cdot 10^3$, C) $S=10^3$ and D) $S=2 \cdot 10^2$. In each panel 5 viscosities are shown, corresponding in order from the top curve to the bottom curve to $R_v=10^4$, $R_v=2 \cdot 10^3$, $R_v=10^3$, $R_v=2 \cdot 10^2$, $R_v=10^2$. Note that in the legends, the viscosity is reported directly rather than the Reynolds number: in our normalized units they are simply one the reciprocal of the other.}\label{confronto_mu_x}
\end{figure}

 Figure \ref{confronto_mu_x} shows the summary of the different runs conducted varying resistivity and viscosity. Each panel corresponds to a different resistivity and reports the time evolution of the maximum difference in the averaged profile $<u_x(z)>_x$, corresponding to the differential between the flank velocity (fast solar wind) and the central plasma (slow solar wind).  In each panel several runs are displayed, corresponding to different viscosities in each run.

Two trends are clear.
First, for some runs, the same progressive acceleration seen in Fig.~\ref{windvst} above is followed by a disruptive phase when the boundary conditions affirm their presence. 
 In some cases at low $R_v$, the system never reaches this state before the completion of the run. 
Second, observing the progressive acceleration phase, one unmistakable conclusion emerges: the speed of acceleration (i.e. the slope of the curve in the figures) is hardly affected by the resistivity  but scales monotonically and markedly with the viscosity. The conclusion regarding the comparatively lower sensitivity of the acceleration mechanism to resistivity was already discussed in a previous work~\citep{lapenta-wind}. Here the focus is on the much more apparent sensitivity to viscosity. As viscosity is increased by two orders of magnitude from a Reynolds number of  $R_v=10^4$ to $R_v=10^2$, the slope increases markedly: for example, the speed at time of $t/\tau_A=200$ for the runs with $S=10^4$ drops by $10 \%$ in the $R_v=10^4$ case and $55 \%$ in the $R_v=10^2$ case. Clearly the mechanism for slow wind acceleration appears to be more linked to the process of viscous drag than to the processes related to resistivity and island formation.

The either/or test just conducted seemingly appears to conclude that the main mechanism in action in the simulations is just simple viscous drag. However, a closer scrutiny reveals that this conclusion based on the overall evolution of the systems still misses one important contribution that the island formation provides. The next subsection  reveals this more subtle effect.

\subsection{Role of the Electric  Field}

A key aspect of the evolution of the system considered above rests in the role of the electric field. 
The initial equilibrium is already characterised by an electrostatic field needed to produce the velocity shear imposed initially. The subsequent evolution of the tearing mode leads to the growth also of an electromagnetic field.

 In the present MHD treatment, the plasma flow is primarily due to the drift motion caused by the presence of electric fields:
\begin{equation}
 {\bf v}_{{\bf E} \times {\bf B}} = \frac{{\bf E} \times {\bf B}}{B^2}
\end{equation}

Once the electric field is written in terms of the vector and scalar potential, the drift clearly shows the presence of a dual nature:
\begin{equation}
 {\bf v}_{{\bf E} \times {\bf B}}=-\frac{1}{B^2}(\nabla \varphi \times {\bf B} +\frac{\partial {\bf A}}{\partial t} \times {\bf B})
\end{equation}
The flow can be of two natures: electrostatic and electromagnetic. The initial configuration has only a $v_x(z)$ component of the flow that is caused  by a $E_z$ component of the electric field: therefore the electric field has zero curl and it is purely electrostatic. The tearing mode is a mode primarily involving the evolution of the out of plane component of the vector potential ${\bf A}$ and it is inductive in nature since it produces an electromagnetic field thanks to the topological variations of B.  However, in presence of out of plane components of the field and of flow shears, there is also an electrostatic potential associate with it~\citep{daughton}.

The viscous effect alone, instead, does not directly affect the magnetic field and causes electrostatic fields. If we consider the evolution of the  initial state under the action of viscosity alone, the full set of equations reduces to just:
\begin{equation}
 \frac{\partial u_x}{\partial t} = -\frac{1}{R_v}\frac{\partial^2u_x}{\partial z} \label{viscosity}
\end{equation}
governing the only remaining variable $u_x(z)$. Using Ohm's law:
\begin{equation}
 \nabla \times E= - {\bf V} \nabla \cdot {\bf B} + {\bf B} \nabla \cdot  {\bf V}
- {\bf B} \cdot \nabla {\bf v} + {\bf v} \cdot \nabla {\bf B}
\end{equation}
where all terms on the right-hand side are zero: the first because of the absence of magnetic mono-poles, the second because the only non-zero component is $u_x$ which is only a function of $z$ (and therefore the divergence of ${\bf u}$ is zero), the third because $u_x$ is constant along field lines, and the last because, similarly, $\bf B$ does not change along the flow lines. The variation of the velocity field generated by viscous drag causes an electric field with zero curl, i.e. it generates an electrostatic field. 

The effect of viscosity alone can be computed easily by solving directly eq.~(\ref{viscosity}). As it is well known~\citep{self-similar}, a linear diffusion equation like eq.~(\ref{viscosity}) admits a self-similar analytical solution of the form: 
\begin{equation}
u_x=u_0\frac{1}{t^{1/2}} e^{-R_v z^2/4t}
\end{equation}
that provides a simple scaling law for the time-variation of the differential in the speed of the central initially stagnant plasma and the outer fast solar wind: $\Delta V_x \propto 1/\sqrt{t}$. 

\begin{figure}[t]
\vspace*{2mm}
\begin{center}
\includegraphics[width =8
cm]{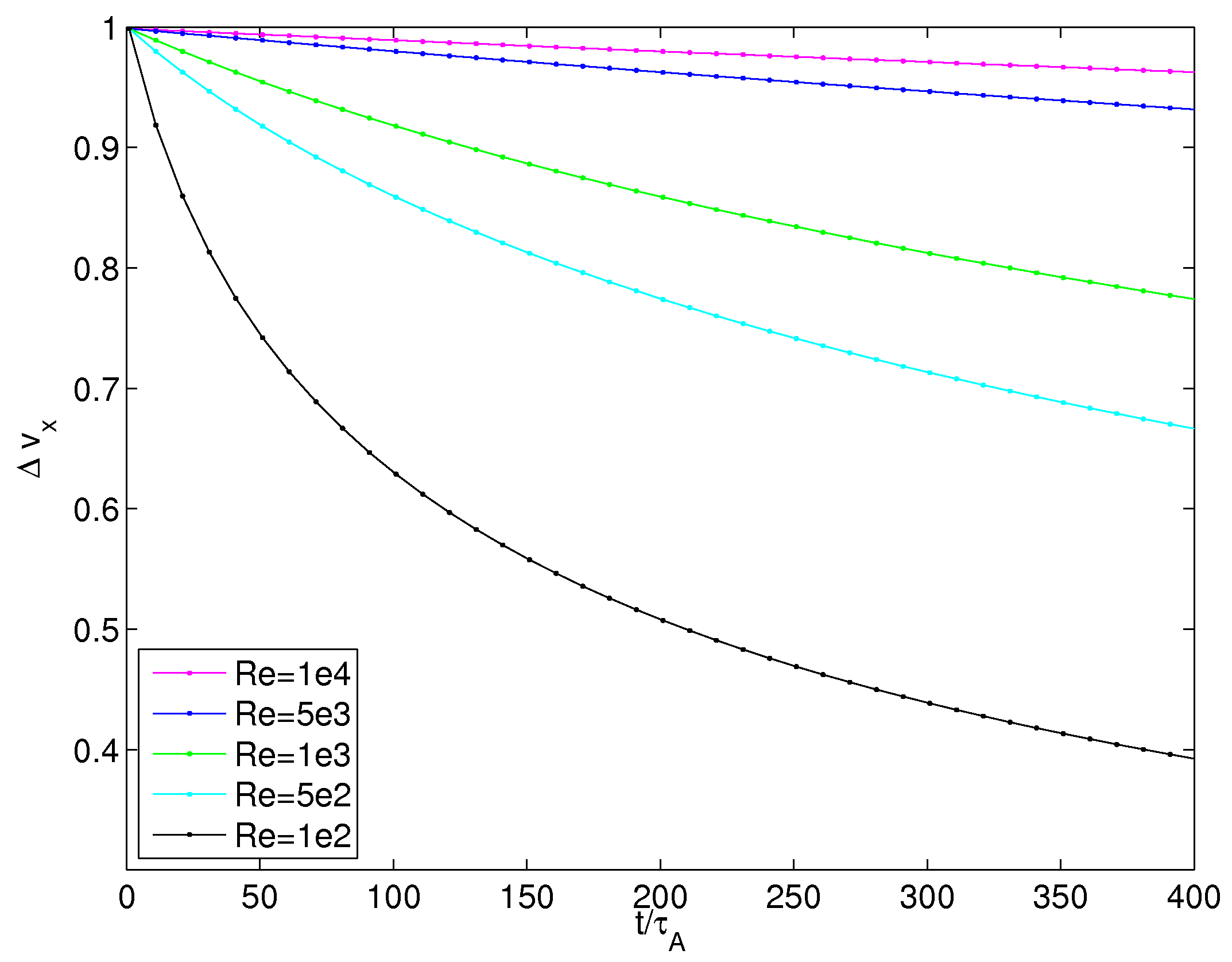}
\end{center}
\caption{Evolution of the difference in velocity $\Delta V_x$ between the central plasma (initially at rest) and the outer fast solar wind. The results are from a simple purely viscous simulation that neglects any magnetic field and resistive effect. Five viscosities are shown, from top to bottom: $R_v=10^4$, $R_v=5 \cdot 10^3$, $R_v=10^3$, $R_v=5 \cdot 10^2$ and $R_v=10^2$.}\label{viscous_drag}
\end{figure}

To illustrate the effect of viscosity alone, we have solved the equation for viscous drag (i.e. eq.~(\ref{viscosity})) numerically starting with the same initial profile considered in the full MHD simulations and for the same ranges of viscosity. Figure \ref{viscous_drag} shows the evolution in time of the jump in speed between the central plasma and the outer fast solar wind. The results can be compared directly with Fig.~\ref{confronto_mu_x} for the full MHD simulations. As can be observed the additional effect of the presence of the magnetic field evolution is very significant. The general shape of the time evolution is the same in the purely viscous case as in the full MHD case, and in both cases resembles the inverse-square-root power law predicted by the self-similar solution. A quantitative comparison reveals that the evolution with full MHD has a very much increased rate of decay. For example, considering the case $R_v=10^3$, at time $t/\tau_A=200$, the full MHD simulation shows a difference in speed of 0.65, while in the purely viscous case this difference is still 0.87. Clearly, the full MHD evolution provides additional drag forces that transcend the direct viscous drag. 

The additional force is coming from the electric field. 

\begin{figure}[t]
\vspace*{2mm}
\begin{center}
\includegraphics[width =8
cm]{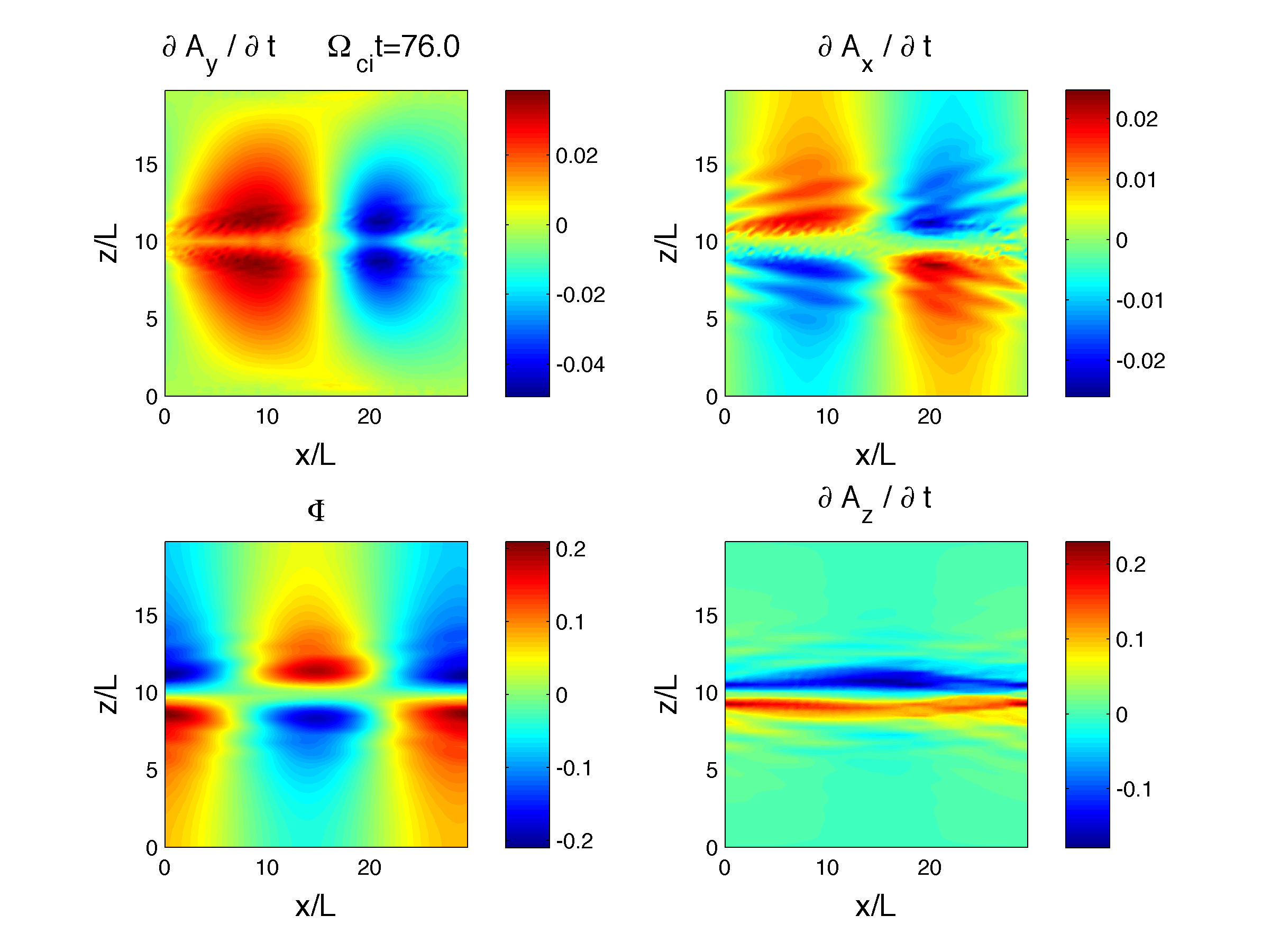}
\end{center}
\caption{ Vector and scalar potentials at time $t=76$ for a run  starting from the  initial force free reversed field equilibrium with $S=10^4$ and $R_v=10^4$ (same run considered in Fig.~\ref{it15}).}\label{potenziale}
\end{figure}

\begin{figure}[t]
\vspace*{2mm}
\begin{center}
\includegraphics[width =8
cm]{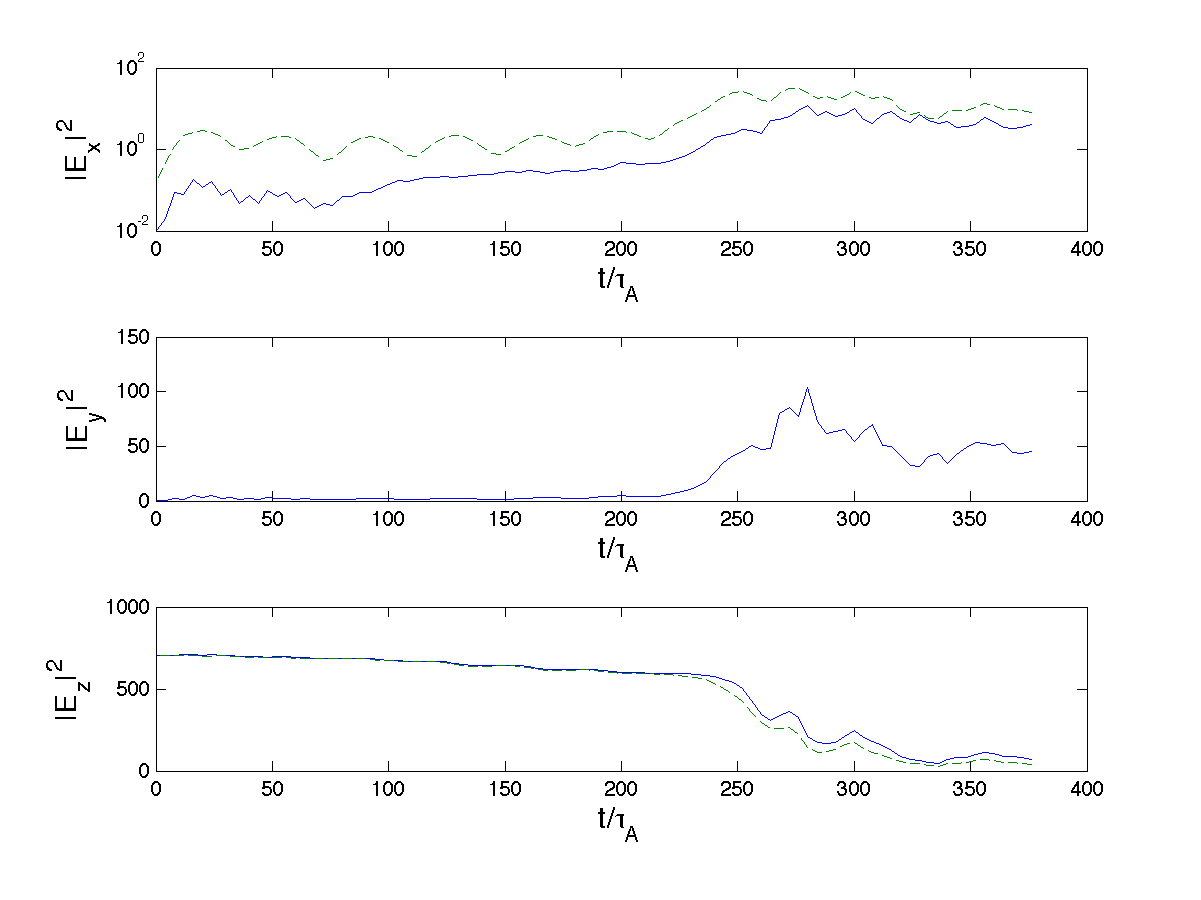}
\end{center}
\caption{ History of the electric field energy for a run with the initial force free reversed field equilibrium and with  $S=10^4$ and $R_v=10^4$. The electrostatic component is shown as a dashed line, the total as a solid line. Note that in a 2D simulation in the $(x,z)$ plane only the two in-plane components can be present for  the electrostatic field. }\label{field-history}
\end{figure}

Figure \ref{potenziale} shows the vector and scalar potential at time $t/\tau_A=76$,  for the same simulation shown in Figs.~\ref{it15} above. A significant electrostatic field is present form the beginning and it is caused, as noted above, by the need to support the initial sheared velocity field. The history of the electric field energy is shown in Fig.~\ref{field-history}. The partial contribution  of the electrostatic field is shown as a dashed line. Except for the $z$ component that includes the initial strong electrostatic field supporting the sheared flow, the electromagnetic component proper of the tearing mode dominates all components.  The electric field perturbation is therefore primarily caused by the tearing growth and not by viscous drag.

The tearing instability developing in the centre of the system moves with respect to the plasma in the  flanks as shown in Fig.~\ref{it15}. The island remain localised in the centre of the plasma and it comoves with the local plasma speed. The direct viscous drag is small. However the island's electric field structure extends far beyond it and well into the flanks, as shown in Fig.~\ref{potenziale}. These fields move with the island speed and have a relative speed with respect to the fast solar wind in the flanks. The expanded range of influence of the island determined by the electric fields allows for an enhanced drag that no longer happens only across layers touching each other but extends globally as the electric field of the island feels the drag force over the whole domain.

\section{Non-linear evolution of the initial 2D helmet streamer equilibrium}

The 2D helmet streamer configuration considered is characterised by the presence of a cusp region above the helmet streamer where the field lines from converging towards the cusp at lower altitudes become essentially parallel. In this region the field-aligned speed also changes direction from a converging motion at lower altitudes to a parallel motion above. In the  1D equilibrium considered above, the focus was entirely on the region above the cusp where the flow and filed lines are initially straight and parallel. The 2D equilibrium with cusp, instead, allows us to consider also the motion in the lower part where the flow and the field lines are converging. 

Converging motions are particularly relevant to the formation of blobs observed in the LASCO images \citep{wang}: converging motions cause magnetic fluxes to pile up towards the converging point and can drive magnetic reconnection. Magnetic reconnection driven by macroscopic flows becomes enslaved to the flows and its rate no longer depends on the details of the microscopic physics allowing reconnection~\citep{biskamp-recon, dorelli,chacon}. 
A previous study showed that in the 2D configuration under consideration the flow towards the cusp produces the conditions for a reconnection process that progresses at the speed dictated by the incoming flux, insensitive to the resistive processes allowing the break up of the frozen-in condition~\citep{lapenta-wind}. As illustrated by \citet{lapenta-wind}, the mechanism allowing the decoupling of resistivity and reconnection rate is the flux pile up: flux piles up in front of the diffusion region to bring the speed of reconnection up to the needed rate~\citep{dorelli}. Indeed, increasing the magnetic field strength at the inflow side of the diffusion region increases the Alfven speed and speeds up the whole process of reconnection. 

In the present study, we consider, instead, the role of viscosity. As the brief summary of the previous understanding implies, the expected outcome is that in driven reconnection viscosity should be a diminishing factor: as viscosity increases, the flows that drive reconnection become impeded by the viscous drag and see their ability to drive reconnection diminished. Therefore, the expectation is that viscosity is not a needed ingredient in the physics leading to  blob formation in the region around the cusp. 

\begin{figure}[t]
\vspace*{2mm}
\begin{center}
A) $R_v=10^3$\\
\includegraphics[width=8
cm]{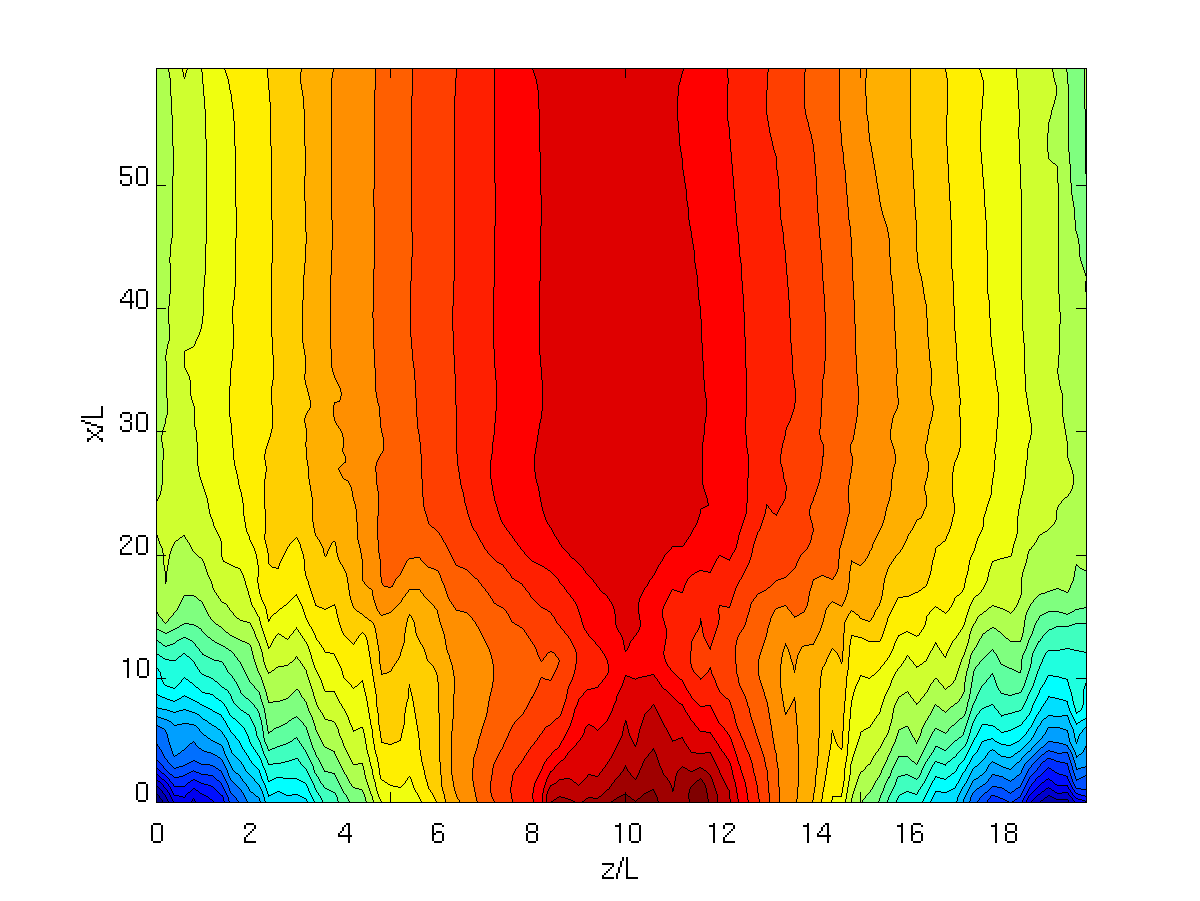}\\
B) $R_v=10^4$\\
\includegraphics[width=8
cm]{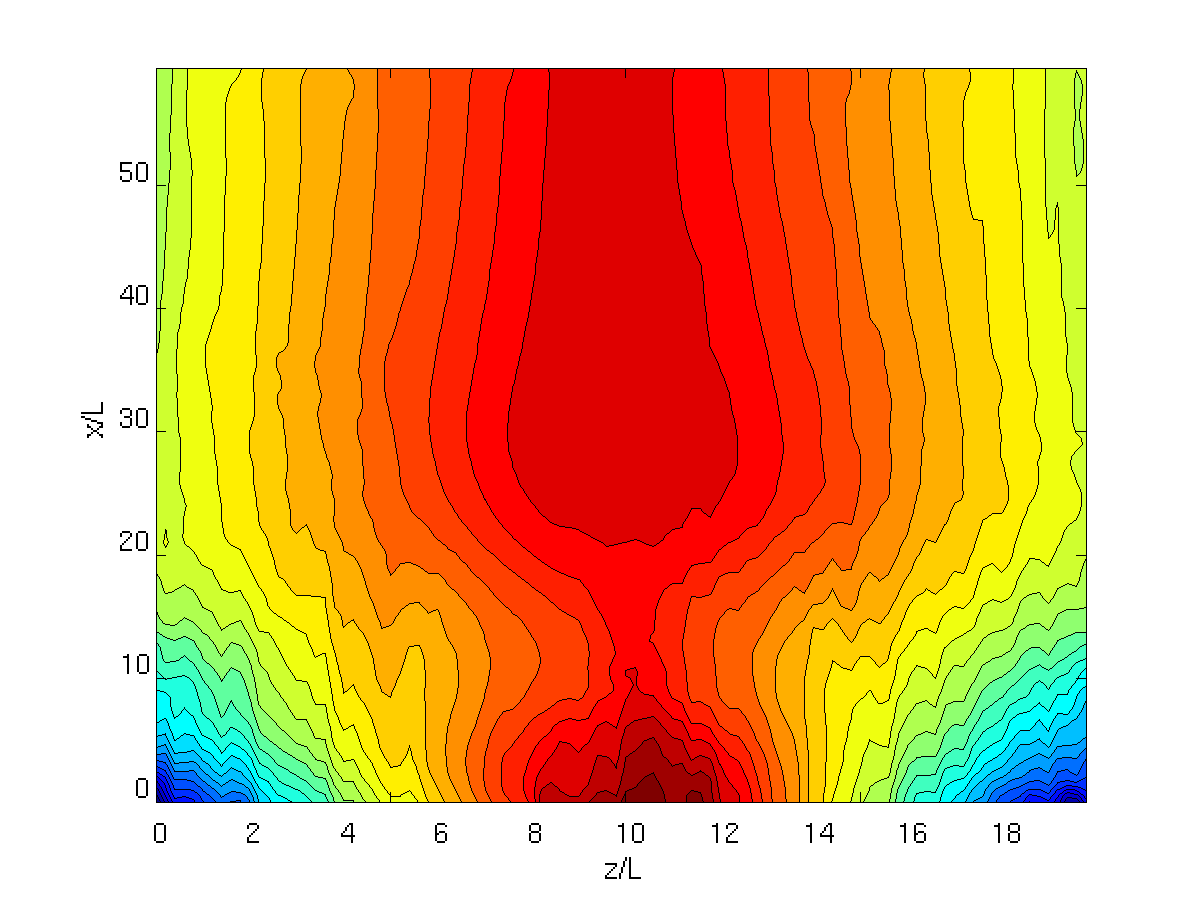}
\end{center}
\caption{Non-linear evolution of the initial 2D helmet streamer equilibrium. Contours of the vector potential $A_y$, representing magnetic surfaces, 
at time $t/\tau_A=200$} \label{cusp_island}
\end{figure}

To investigate this point, Fig.~\ref{cusp_island} shows the magnetic surfaces after the formation of a magnetic island above the cusp of the helmet streamer. As can be observed, indeed the process happens regardless of viscosity, and, as expected, the more viscous run shows a diminished speed of formation of the island.  At a given same time in the two runs, the island is less developed in the higher viscosity case.

\begin{figure}[t]
\vspace*{2mm}
\begin{center}
\includegraphics[width=8
cm]{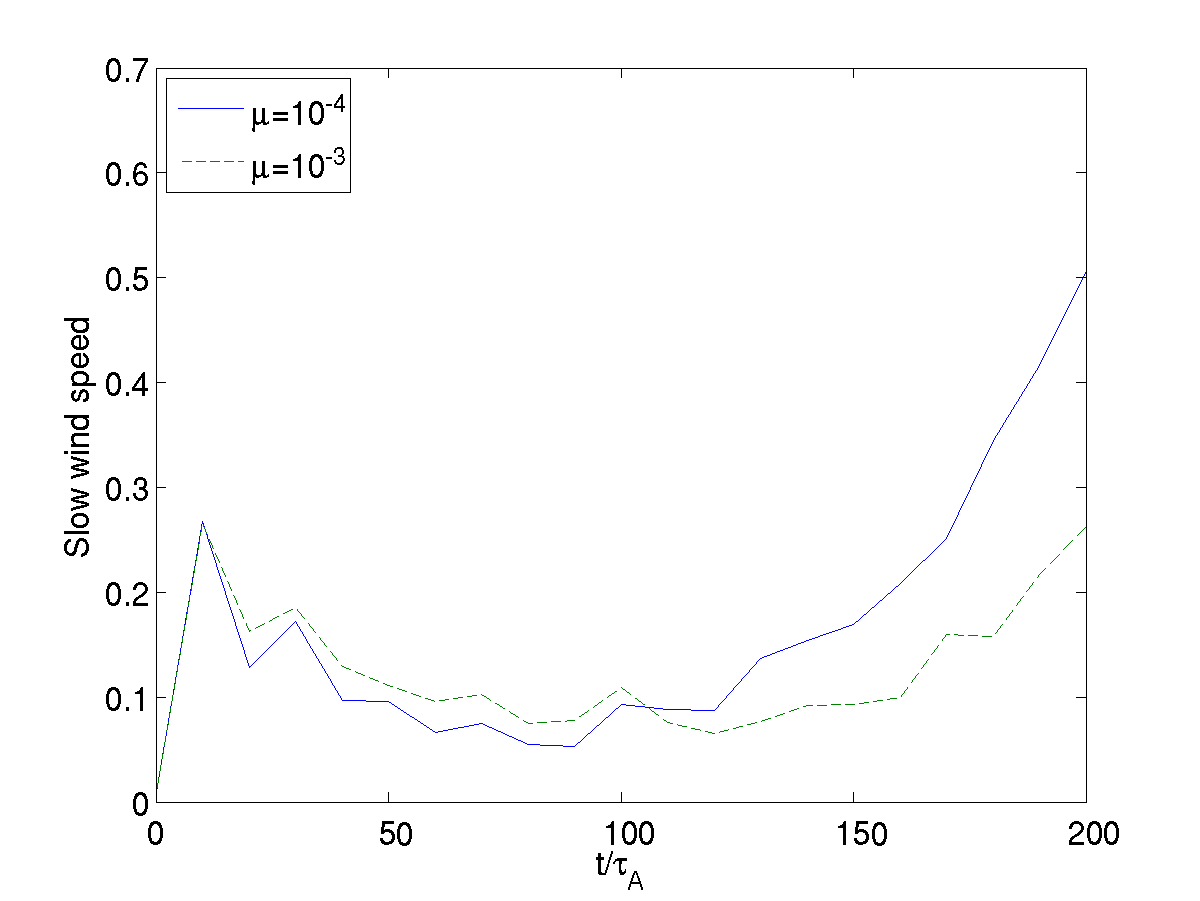}
\end{center}
\caption{Non-linear evolution of the initial 2D helmet streamer equilibrium. Comparison of the time evolution for the reconnected flux in two 
runs with different viscosities ($R_v=10^3$ and $R_v=10^4$) but equal resistivity
$S=10^4$.} \label{cusp_rec}
\end{figure}

This conclusion is further confirmed by Fig.~\ref{cusp_rec} that shows the evolution of the reconnected flux for the two runs with different viscosity (but equal resistivity) The reconnection process is similar but slightly faster for the lower viscosity run. 

However, the fundamental difference of the 2D equilibrium when compared with the 1D equilibrium of the previous section is that also the acceleration of the blob is independent of viscosity. The mechanism allowing the acceleration is no longer the viscous drag exerted by the outer plasma on the forming island (and mediated by the electric field) but rather is directly the converging flow at the cusp. 

\begin{figure}[t]
\vspace*{2mm}
\begin{center}
A) $R_v=10^3$\\
\includegraphics[width=8
cm]{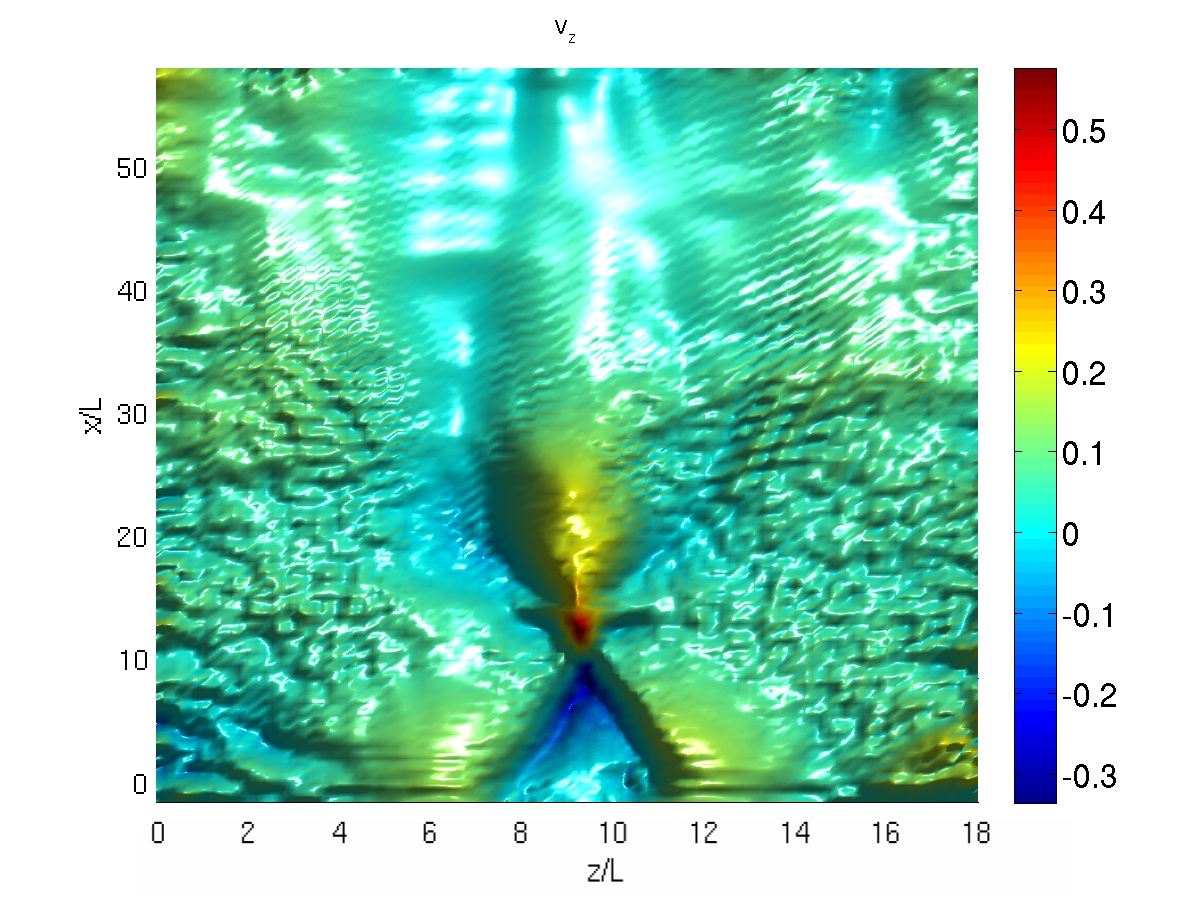}\\
B) $R_v=10^4$\\
\includegraphics[width=8
cm]{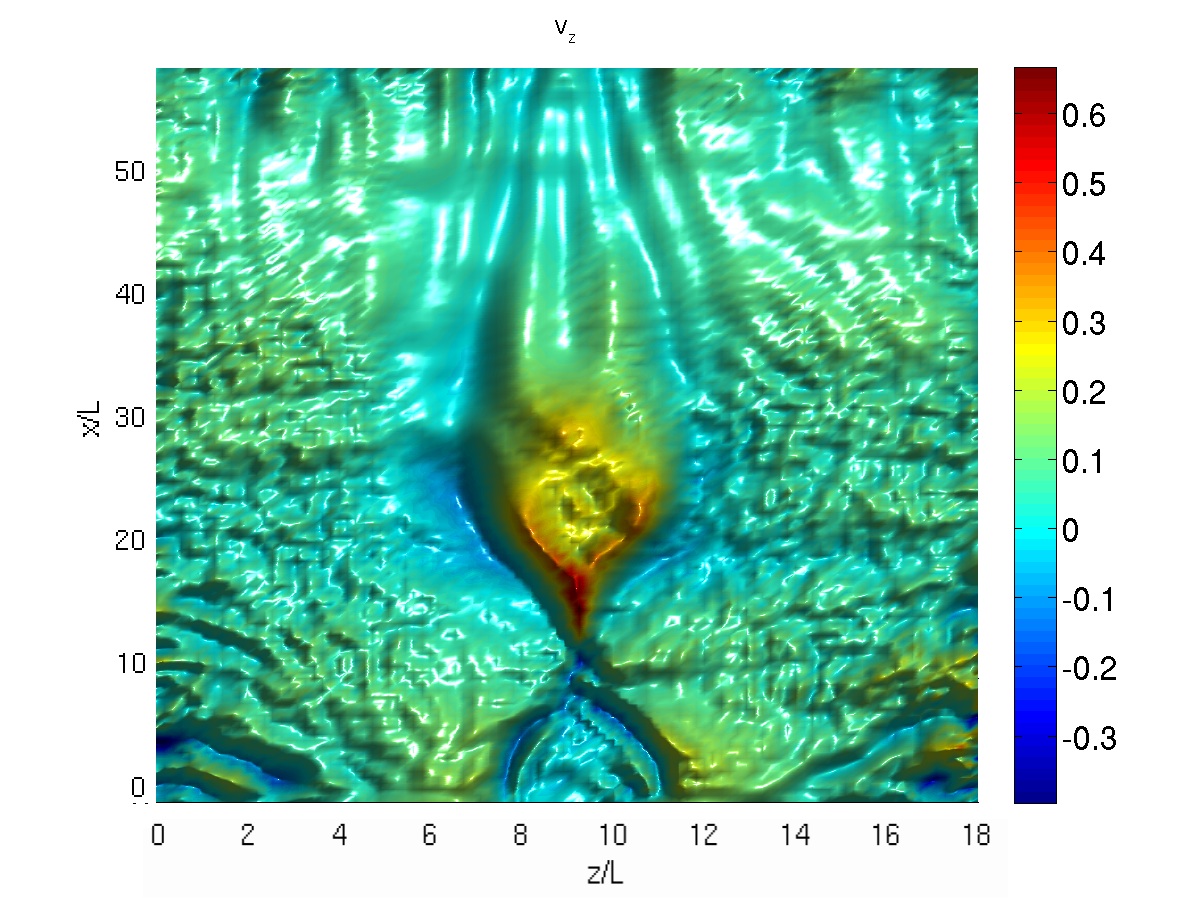}
\end{center}
\caption{Non-linear evolution of the initial 2D helmet streamer equilibrium. False colour representation of the vertical speed away from the Sun $u_x$, 
at time $t/\tau_A=200$. Two runs are shown with different viscosities (panel A:$R_v=10^3$ and panel B:  $R_v=10^4$) but equal resistivity
$S=10^4$.} \label{cusp_vel}
\end{figure}

Figure \ref{cusp_vel} shows the vertical speed in the same two runs with different viscosity considered above. The flow velocity originating from the open field lines converges and focuses at the cusp in the region right below the forming island. The island finds itself in the same situation of a ball sitting atop a jet in a fountain: it is pushed by the jetting plasma. The converging flow focuses at the cusp and pushes the island upward by direct momentum transfer from the plasma to the island field lines. Viscous drag is no longer needed. 

\begin{figure}[t]
\vspace*{2mm}
\begin{center}
\includegraphics[width=8
cm]{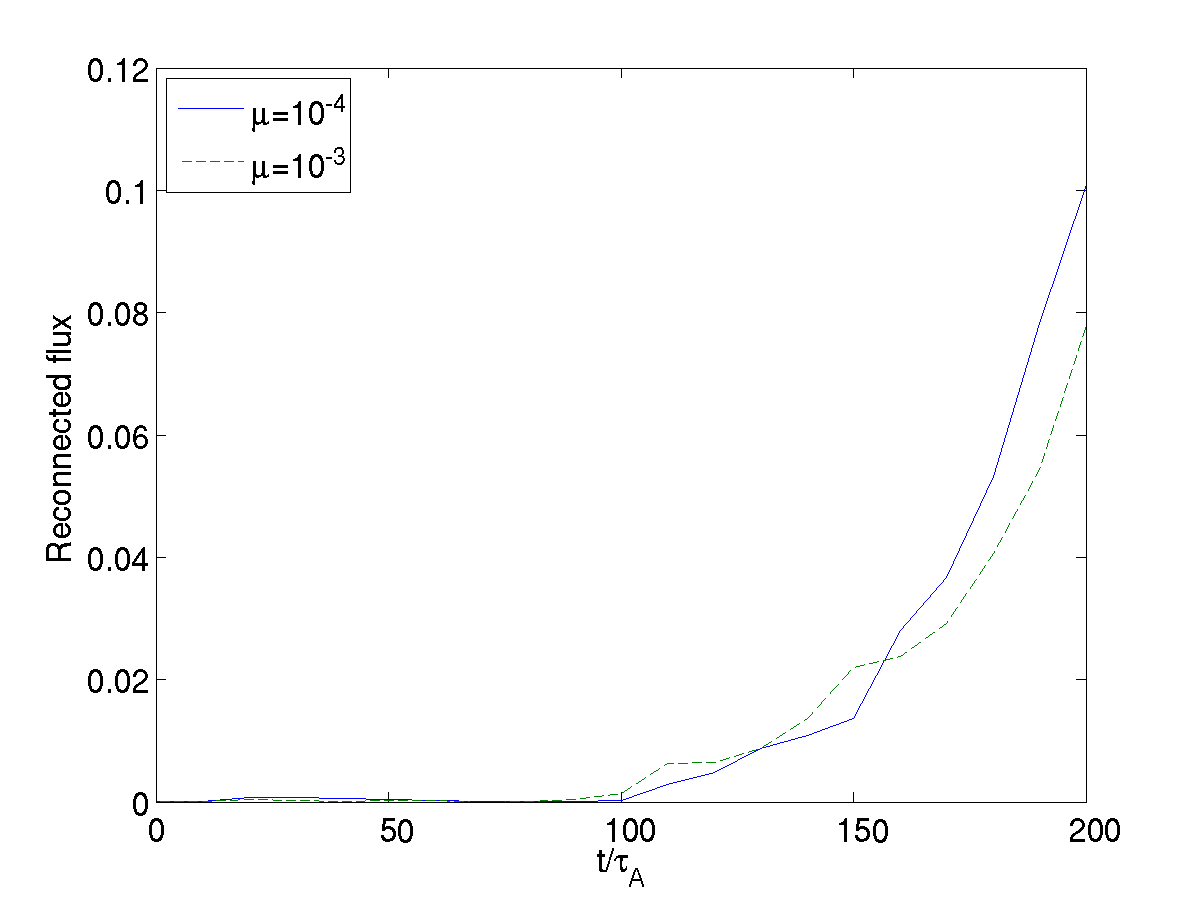}
\end{center}
\caption{Comparison of the time evolution of the vertical velocity in the central axis in two 
runs starting from the  initial 2D helmet streamer equilibrium for the maximum with different viscosities ($R_v=10^3$ and $R_v=10^4$) but equal resistivity
$S=10^4$.} \label{cusp_wind}
\end{figure}

This last conclusion is further confirmed by Fig.~\ref{cusp_wind} that shows the acceleration of the initially stagnant plasma above the cusp as a function of time. Clearly, viscosity is not the cause of the acceleration but rather just as it slows down by a small but perceptible amount the process of reconnection, it also slows down the ability of the converging flow to focus at the cusp and push the forming island upward. 

\conclusions

We have investigated the role of viscosity in the physical mechanisms responsible for the formation of blobs in the slow solar wind. We started from two types of models. The first is a initial 1D equilibrium proposed by \citet{einaudi} and designed to capture the field reversal and the flow shear present in the plasma above the helmet streamers in the regions where blobs have been observed to form in the LASCO images~\citep{wang}. The second model is based on a 2D equilibrium that extends the previous case by including the effect of converging flows at the cusp immediately above a helmet streamer~\citep{lapenta-wind}. The role of viscosity is different in the two cases. 

In the 1D initial equilibrium, viscosity is shown to provide the basic mechanism for momentum transfer from the slow solar wind to the forming blob. The viscous drag produces a progressive acceleration of the plasma blob. The rate of increase of the slow solar wind speed (acceleration) is proportional to the viscosity in the simulation. However, an unexpected finding is that viscosity is not acting alone in its usual local drag action across layers of flowing plasma. Rather, the coupling between the accelerating island and the fast solar wind in the flanks is mediated by the global electric fields caused by the tearing instability forming the island. A direct numerical comparison of the evolution under pure viscous action and the full MHD evolution including the generation of electromagnetic fields due to the tearing instability proves that the drag action is much increased from simple viscous drag.

In the 2D initial configuration, the momentum transfer action is completely different. The viscous drag is not a determining factor in that it does not provide the mechanism for momentum transfer from the fast wind to the forming island. In the 2D case, the flow merges at the cusp and simply pushes the island by direct momentum transfer acting in the same way a ball is pushed up by jetting water in a fountain. The momentum transfer in this case is coming directly by the action of plasma particles directed toward the magnetic filed lines of the island and being deflected by them. Their change in momentum of deflected particles provides a direct accelerating force. Viscosity has no role to play and its effect is just to diminish the flows and to reduce the rate of island formation and its subsequent acceleration. 

The results presented are relevant to the Sun where viscosity and resistivity due to direct collisional transport are extremely low. The ability to explain both in the 1D and in the 2D case how momentum transfer is provided by extra effects not directly due to viscosity is crucial. The results obtained here, as all other results obtainable within the range of currently achievable simulation cannot properly use the viscosity and resistivity of the real corona and cannot account self-consistently for anomalous processes. It is therefore crucial to show that the accelerating mechanisms presented here are indeed stronger than purely viscous processes would produce and can sustain the scaling to realistic coronal parameters.

We remind the reader that we conducted the research here on a Cartesian geometry where the melon seed effect caused by the expanding geometry around the Sun is not considered~\citep{rappazzo}: the results of the present work need to be integrated by considering that additional effect. Future work will consider how the conclusions reached here are modified by additional geometric effects.

\begin{acknowledgements}
Fruitful discussions with Russell Dahlburg and Dana Knoll are gratefully acknowledged. The present work is supported by the {\it Onderzoekfonds K.U. Leuven} (Research Fund KU Leuven), by the European Commission through the SOLAIRE network (MRTN-CT-2006-035484), by the NASA Sun Earth Connection Theory Program and by
 the LDRD program at the
Los Alamos National Laboratory. Work performed in part under the auspices of
the National Nuclear Security Administration of the U.S. Department
of Energy by the Los Alamos National Laboratory, operated by Los
Alamos National Security LLC under contract DE-AC52-06NA25396.
Simulations conducted in part on the
HPC cluster VIC of the Katholieke Universiteit Leuven.

\end{acknowledgements}


\bibliographystyle{egu}
\bibliography{angeo-2007-0074}

\begin{thebibliography}{26}
\providecommand{\natexlab}[1]{#1}
\providecommand{\url}[1]{{\tt #1}}
\providecommand{\urlprefix}{}
\expandafter\ifx\csname urlstyle\endcsname\relax
  \providecommand{\doi}[1]{doi:\discretionary{}{}{}#1}\else
  \providecommand{\doi}{doi:\discretionary{}{}{}\begingroup
  \urlstyle{rm}\Url}\fi

\bibitem[{{Barenblatt}(1996)}]{self-similar}
{Barenblatt}, G.~I.: {Scaling, Self-similarity, and Intermediate Asymptotics},
  Cambridge University Press, Cambridge, 1996.

\bibitem[{{Bettarini} et~al.(2006){Bettarini}, {Landi}, {Rappazzo}, {Velli},
  and {Opher}}]{bettarini}
{Bettarini}, L., {Landi}, S., {Rappazzo}, F.~A., {Velli}, M., and {Opher}, M.:
  {Tearing and Kelvin-Helmholtz instabilities in the heliospheric plasma},
  Astron. Astrophys., 452, 321--330, 2006.

\bibitem[{{Biskamp}(1997)}]{biskamp}
{Biskamp}, D.: Nonlinear Magnetohydrodynamics, Cambridge University Press,
  Cambridge, UK, 1997.

\bibitem[{{Biskamp}(2000)}]{biskamp-recon}
{Biskamp}, D.: {Magnetic Reconnection in Plasmas}, Cambridge University Press,
  Cambridge, UK, 2000.

\bibitem[{{Brackbill}(1991)}]{brackbill}
{Brackbill}, J.~U.: {FLIP MHD - A particle-in-cell method for
  magnetohydrodynamics}, Journal of Computational Physics, 96, 163--192, 1991.

\bibitem[{Braginskii(1965)}]{braginskii}
Braginskii, S.~I.: Rev. Plasma Physics (I), p. 205, Consultants Bureau
  Enterprises, New York, 1965.

\bibitem[{{Daughton} and {Karimabadi}(2005)}]{daughton}
{Daughton}, W. and {Karimabadi}, H.: {Kinetic theory of collisionless tearing
  at the magnetopause}, J. Geophys. Res., 110, 3217, 2005.

\bibitem[{{Dorelli} and {Birn}(2003)}]{dorelli}
{Dorelli}, J.~C. and {Birn}, J.: {Whistler-mediated magnetic reconnection in
  large systems: Magnetic flux pileup and the formation of thin current
  sheets}, J. Geophys. Res., 108, 1133, 2003.

\bibitem[{Einaudi et~al.(1999)Einaudi, Bonicelli, Dahlburg, and
  Karpen}]{einaudijgr}
Einaudi, G., Bonicelli, P., Dahlburg, R., and Karpen, J.: {Formation of the
  slow solar wind in a coronal streamer}, J. Geophys. Res., 104, 521, 1999.

\bibitem[{{Einaudi} et~al.(2001){Einaudi}, {Chibbaro}, {Dahlburg}, and
  {Velli}}]{einaudi}
{Einaudi}, G., {Chibbaro}, S., {Dahlburg}, R.~B., and {Velli}, M.: {Plasmoid
  Formation and Acceleration in the Solar Streamer Belt}, Astrophys. J., 547,
  1167--1177, \doi{10.1086/318400}, 2001.

\bibitem[{{Endeve} et~al.(2003){Endeve}, {Leer}, and {Holzer}}]{endeve}
{Endeve}, E., {Leer}, E., and {Holzer}, T.~E.: {Two-dimensional
  Magnetohydrodynamic Models of the Solar Corona: Mass Loss from the Streamer
  Belt}, Astrophys. J., 589, 1040--1053, 2003.

\bibitem[{{Fisk} and {Schwadron}(2001)}]{fisk}
{Fisk}, L.~A. and {Schwadron}, N.~A.: {Origin of the Solar Wind: Theory}, Space
  Sci. Rev., 97, 21--33, 2001.

\bibitem[{Goedbloed and Poedts(2004)}]{poedts}
Goedbloed, J. P.~H. and Poedts, S.: {Principles of Magnetohydrodynamics},
  Cambridge University Press, Cambridge, 2004.

\bibitem[{{Goossens}(2003)}]{goossens}
{Goossens}, M.: {An introduction to plasma astrophysics and
  magnetohydrodynamics}, Kluwer, Dordrecht, The Nederlands, 2003.

\bibitem[{{Knoll} and {Chac{\'o}n}(2002)}]{chacon}
{Knoll}, D.~A. and {Chac{\'o}n}, L.: {Magnetic Reconnection in the
  Two-Dimensional Kelvin-Helmholtz Instability}, Physical Review Letters, 88,
  215\,003, 2002.

\bibitem[{Landau and Lifshitz(1959)}]{landau}
Landau, L. and Lifshitz, E.: Fluid Mechanics, vol.~6 of {\em Course of
  Theoretical Physics\/}, Pergamon, London, U.K., 1959.

\bibitem[{{Lapenta} and {Knoll}(2003)}]{lapenta-solar}
{Lapenta}, G. and {Knoll}, D.~A.: {Reconnection in the Solar Corona: Role of
  the Kelvin-Helmholtz Instability}, Solar Phys., 214, 107--129, 2003.

\bibitem[{{Lapenta} and {Knoll}(2005)}]{lapenta-wind}
{Lapenta}, G. and {Knoll}, D.~A.: {Effect of a Converging Flow at the Streamer
  Cusp on the Genesis of the Slow Solar Wind}, Astrophys. J., 624, 1049--1056,
  \doi{10.1086/429262}, 2005.

\bibitem[{{Rappazzo} et~al.(2005){Rappazzo}, {Velli}, {Einaudi}, and
  {Dahlburg}}]{rappazzo}
{Rappazzo}, A.~F., {Velli}, M., {Einaudi}, G., and {Dahlburg}, R.~B.:
  {Diamagnetic and Expansion Effects on the Observable Properties of the Slow
  Solar Wind in a Coronal Streamer}, Astrophys. J., 633, 474--488, 2005.

\bibitem[{{Schmidt} and {Cargill}(2000)}]{schmidt}
{Schmidt}, J.~M. and {Cargill}, P.~J.: {A model for accelerated density
  enhancements emerging from coronal streamers in Large-Angle and Spectrometric
  Coronagraph observations}, J. Geophys. Res., 105, 10\,455--10\,464, 2000.

\bibitem[{{Sheeley} et~al.(1997){Sheeley}, {Wang}, {Hawley}, {Brueckner},
  {Dere}, {Howard}, {Koomen}, {Korendyke}, {Michels}, {Paswaters}, {Socker},
  {St.~Cyr}, {Wang}, {Lamy}, {Llebaria}, {Schwenn}, {Simnett}, {Plunkett}, and
  {Biesecker}}]{sheeley}
{Sheeley}, Jr., N.~R., {Wang}, Y.-M., {Hawley}, S.~H., {Brueckner}, G.~E.,
  {Dere}, K.~P., {Howard}, R.~A., {Koomen}, M.~J., {Korendyke}, C.~M.,
  {Michels}, D.~J., {Paswaters}, S.~E., {Socker}, D.~G., {St.~Cyr}, O.~C.,
  {Wang}, D., {Lamy}, P.~L., {Llebaria}, A., {Schwenn}, R., {Simnett}, G.~M.,
  {Plunkett}, S., and {Biesecker}, D.~A.: {Measurements of Flow Speeds in the
  Corona between 2 and 30 R sub sun}, Astrophys. J., 484, 472, 1997.

\bibitem[{{van Aalst} et~al.(1999){van Aalst}, {Martens}, and
  {Beli{\"e}n}}]{vanaalst}
{van Aalst}, M.~K., {Martens}, P.~C.~H., and {Beli{\"e}n}, A.~J.~C.: {Can
  Streamer Blobs Prevent the Buildup of the Interplanetary Magnetic Field?},
  Astrophys. J. Lett., 511, L125--L128, 1999.

\bibitem[{{Wang} et~al.(1998){Wang}, {Sheeley}, {Walters}, {Brueckner},
  {Howard}, {Michels}, {Lamy}, {Schwenn}, and {Simnett}}]{wang}
{Wang}, Y.-M., {Sheeley}, Jr., N.~R., {Walters}, J.~H., {Brueckner}, G.~E.,
  {Howard}, R.~A., {Michels}, D.~J., {Lamy}, P.~L., {Schwenn}, R., and
  {Simnett}, G.~M.: {Origin of Streamer Material in the Outer Corona},
  Astrophys. J. Lett., 498, L165, 1998.

\bibitem[{{Wiegelmann} et~al.(1998){Wiegelmann}, {Schindler}, and
  {Neukirch}}]{schindler}
{Wiegelmann}, T., {Schindler}, K., and {Neukirch}, T.: {Helmet Streamers with
  Triple Structures: Weakly Two-Dimensional Stationary States}, Solar Phys.,
  180, 439--460, 1998.

\bibitem[{{Wiegelmann} et~al.(2000){Wiegelmann}, {Schindler}, and
  {Neukirch}}]{wiegelmann}
{Wiegelmann}, T., {Schindler}, K., and {Neukirch}, T.: {Helmet Streamers with
  Triple Structures: Simulations of resistive dynamics}, Solar Phys., 191,
  391--407, 2000.

\bibitem[{{Wu} et~al.(2000){Wu}, {Wang}, {Plunkett}, and {Michels}}]{wu}
{Wu}, S.~T., {Wang}, A.~H., {Plunkett}, S.~P., and {Michels}, D.~J.: {Evolution
  of Global-Scale Coronal Magnetic Field due to Magnetic Reconnection: The
  Formation of the Observed Blob Motion in the Coronal Streamer Belt},
  Astrophys. J., 545, 1101--1115, 2000.

\end{thebibliography}

\end{document}